\documentclass[twocolumn,superscriptaddress,floatfix,preprintnumbers,prd,nofootinbib,hyperref]{revtex4-1}

\pdfoutput=1

\usepackage{graphicx,color}
\usepackage{amsmath,amssymb,eqnarray}
\usepackage{hyperref} 

\hypersetup{
    colorlinks=true,
    citecolor=[rgb]{.1, .7, .6},
    linkcolor=[rgb]{.2, .55, .95},
    filecolor=magenta,
    urlcolor=[rgb]{.1, .7, .6},
}

\renewcommand{\vec}[1]{\boldsymbol{#1}}

\usepackage{braket}
\usepackage{tensor}
\begin{document}

\title{Resonant Conversion of Gravitational Waves in Neutron Star Magnetospheres}

\author{Jamie I.~McDonald}
\affiliation{Department of Physics and Astronomy, University of Manchester, Manchester M13 9PL, UK}

\author{Sebastian~A.~R.~Ellis}
\affiliation{D{\'e}partment de Physique Th{\'e}orique, Universit{\'e}  de Gen{\`e}ve,
24 quai Ernest Ansermet, 1211 Gen{\`e}ve 4, Switzerland}

\begin{abstract}
High frequency gravitational waves are the subject of rapidly growing interest in the theoretical and experimental community. In this work we 
calculate the resonant conversion of gravitational waves into photons in the magnetospheres of neutron stars via the inverse Gertsenshtein mechanism. The resonance occurs in regions where the vacuum birefringence effects cancel the classical plasma contribution to the photon dispersion relation,
leading to a massless photon in the medium which becomes kinematically matched to the graviton. We set limits on the amplitude of a possible stochastic background of gravitational waves using X-ray and IR flux measurements of neutron stars.  Using Chandra ($2-8\,\text{keV}$) and NuSTAR ($3-79\,\text{keV}$) observations of RX J1856.6-3754, we set strain limits $h_c^{\rm lim} \simeq 10^{-26} -    10^{-24}$ in the frequency range $ 5\times 10^{17}\, {\rm Hz} \lesssim f \lesssim  2\times 10^{19}\,\text{Hz}$. Our limits are many orders of magnitude stronger than existing constrains from individual neutron stars at the same frequencies. 
We also 
use recent JWST observations of the Magnetar 4U 0142+61
in the range $2.7\times 10^{13}\, {\rm Hz} \lesssim f \lesssim 5.9\times 10^{13}\, {\rm Hz} $,  setting a limit $h_{\rm c}^{\rm lim} \simeq 5 \times 10^{-19}$. These constraints are in complementary frequency ranges to laboratory searches with CAST, OSQAR and ALPS II. We expect these limits to be improved both in reach and breadth with a more exhaustive use of telescope data across the full spectrum of frequencies and targets.

\end{abstract}

\maketitle

\section{Introduction}

The detection~\cite{LIGOScientific:2016aoc} of gravitational waves by LIGO has revolutionised our ability to study the Cosmos and opened the door to gravitational wave astronomy. LIGO \cite{LIGOScientific:2016emj} has since been joined by the VIRGO \cite{VIRGO:2014yos} and KAGRA \cite{Aso:2013eba} detectors to form the LVK collaboration which has since measured hundreds more black hole and neutron star mergers at frequencies of $\mathcal{O}(100 {\rm Hz})$.   More recently, evidence for nHz gravitational waves has emerged from pulsar timing measurements by the NANOGrav, PPTA, EPTA, and InPTA ~\cite{NANOGrav:2020bcs,Goncharov:2021oub,EPTA:2021crs,Tarafdar:2022toa} collaborations. In addition, there remain a number of future detectors aiming to expand the frequency coverage of gravitational wave astronomy, including space-based interferometers like LISA \cite{LISA:2017pwj} ($10^{-2} \rm{Hz}$)  and atom interferometers like MAGIS/AION \cite{Badurina:2019hst,MAGIS-100:2021etm} ($10^{-1}{\rm Hz} - 10 {\rm Hz}$). These efforts mark the arrival of the era of multi-wavelength gravitational wave astronomy. 

This mission echoes the evolution of conventional electromagnetic astronomy, which has proved to be one of the main driving forces behind fundamental physics, from early observations of dark matter \cite{Rubin:1970zza} and dark energy \cite{SupernovaCosmologyProject:1998vns,SupernovaSearchTeam:1998fmf} to the discovery of the cosmic microwave background \cite{Penzias:1965wn}. Today, photon-based astronomy spans 15 orders of magnitude in frequency, with an array of sophisticated telescopes from the radio through to gamma-rays and continues to play an active role in guiding our understanding of the Universe at the most fundamental level. It is worth mentioning that several of these discoveries have been made serendipitously. By analogy, it seems highly likely that gravitational wave astronomy will 
unlock yet more secrets of our Universe as we
explore the full range of the gravitational wave spectrum. 

In recent years, there has been growing interest in pushing gravitational wave astronomy to even higher frequencies above ${\rm kHz}$, (see Ref.~\cite{Aggarwal:2020olq} for a review) paving the way for the study of ultra-high-frequency gravitational waves (UHFGWs). At these high frequencies, astrophysical sources of GWs from within the Standard Model are not expected. However, stochastic backgrounds of UHFGWs can be generated by a variety of sources including cosmic strings (see, e.g., Ref.~\cite{Servant:2023tua}), phase transitions (see, e.g., Ref.~\cite{Hindmarsh:2017gnf}), and even the cosmic microwave gravitational wave background itself~\cite{Ghiglieri:2015nfa,Ringwald:2020ist}. Similarly, transient signals can be generated by primordial black hole mergers~\cite{Hawking:1971ei,Khlopov:2008qy,Carr:2020gox,Carr:2021bzv,Franciolini:2022htd} and the depletion of superradiant bosons in the vicinity of black holes~\cite{Ternov:1978gq,Zouros:1979iw,Detweiler:1980uk,Arvanitaki:2009fg,Arvanitaki:2010sy,Yoshino:2013ofa,Brito:2014wla,Arvanitaki:2014wva,Brito:2015oca,Zhu:2020tht}. Most of these proposed sources result from physics beyond the standard model. Searching for UHFGWs therefore offers an exciting opportunity to probe new physics. 
 
The study of UHFGWs is in its infancy, and it is possible that there are further \textit{Standard Model} sources yet to be predicted by theorists.
In the last few years, a range of novel astrophysical effects have been shown to predict the production of large numbers of light particles~\cite{Prabhu:2021zve,Noordhuis:2022ljw,Witte:2024akb}, illustrating the continuing ability of astrophysical environments to surprise us. Early theoretical studies \cite{Casalderrey-Solana:2022rrn} also suggest MHz gravitational waves may be produced in neutron star mergers due to Quantum Chromodynamics effects. Further Standard Model sources of UHFGWs, if predicted, would offer important milestones for detection,
requiring a wide array of experimental approaches.

Significant progress has recently been made in the study of experimental signatures of UHFGWs thanks to
the techniques which have been developed to study light particles, such as axions. Indeed, these are often directly applicable to laboratory searches of UHFGWs across a range of frequencies, as explored in \cite{Berlin:2021txa,Domcke:2022rgu,Domcke:2023bat}. 
Furthermore,
the emergence of new technologies including
levitated sensors \cite{Arvanitaki:2012cn,Aggarwal:2020umq}, bulk acoustic wave resonantors (BAWs) \cite{Goryachev:2014yra}, high-precision atomic clocks \cite{Bringmann:2023gba} and superconducting cavities \cite{Berlin:2023grv} is making the detection of UHFGWs a tangible possibility.

Extending the light particle analogy further, there has been a thriving symbiosis between laboratory searches and astrophysical probes of light fields~\cite{Baryakhtar:2022hbu}. In particular, stars have proved powerful tools for searching for light particles, and they continue to provide some of the leading constraints relative to laboratory searches in some frequency ranges. Neutron stars, in particular, have been used extensively to study ultra-light particles, including axions in both the radio~\cite{Pshirkov:2007st,Hook:2018iia,Huang:2018lxq,Battye:2019aco,Darling:2020uyo,Battye:2021yue,Darling:2020plz,Foster:2022fxn,Battye:2023oac} and X-ray bands~\cite{Buschmann:2019pfp}. These studies exploit an enhanced coupling between axions and photons both due to the very large magnetic fields of neutron stars, and, in the case of dark matter, the kinematic enhancement of the production process due to the axion and photon dispersion relations becoming degenerate in the neutron star plasma, leading to \textit{resonant} production of photons from light particles. 

A range of sophisticated techniques have now been developed for computing light particle signatures from neutron stars, including numerical modelling of photon transport \cite{Leroy:2019ghm,Witte:2021arp,Battye:2021xvt,McDonald:2023shx,Tjemsland:2023vvc} and analytic \cite{Battye:2019aco,McDonald:2023ohd} as well as numerical \cite{Gines:2024ekm} calculations of the 3D photon production process itself. In this work, we demonstrate how this same resonant production process is also present for UHFGWs propagating through neutron star magnetospheres. 
To date, there have been a few schematic studies of \textit{non-resonant} conversion  \cite{Ito:2023fcr,Dandoy:2024oqg} and neither of these have made use of the latest techniques developed above. We therefore carry out the first treatment of the resonant production process, implementing the state-of-the-art treatments detailed above. 

The structure of this paper is as follows. In Sec.~\ref{sec:Production} we outline the production process of photons from gravitons by adapting the latest techniques developed in \cite{McDonald:2023ohd}. In Sec.~\ref{sec:Neutorn Stars} we outline our model for neutron star magnetospheres and describe the characteristic size of signals. In Sec.~\ref{sec:Constraints} we provide our constraints before finally offering our conclusions in Sec.~\ref{sec:conclusions}.

\section{Resonant Photon Production}\label{sec:Production}

Since we shall work across multiple observational bands where
wavelength, frequency and energy are variously used to label photons, it is useful at this stage to lay out our notation and conventions. We work in natural units in which $\hbar = c = 1$ so that the photon energy, $E_\gamma$, angular frequency $\omega$, frequency $f$ and wavelength $\lambda$ are related by
\begin{equation}
    E_\gamma  = \, \omega =  2 \pi f  = 2\pi/\lambda \ .
\end{equation} 

The interaction between gravitons and photons is captured by the the minimal coupling of the spacetime metric $g_{\mu \nu}$ to electromagnetism, represented by the following action
\begin{equation}
    \mathcal{S} = \int d^4 x \sqrt{-g} \, \left[ \frac{m_p^2}{2} \mathcal{R} - \frac{1}{4} g_{\mu \nu } g_{\rho \sigma} F^{\mu \nu} F^{\rho \sigma} \right],
\end{equation}
where $m_p = 1/\sqrt{8 \pi G}$ is the reduced Planck mass
and $F_{\mu \nu} = \partial_\mu A_\nu - \partial_\nu A_\mu$ is the photon field-strength tensor.  Working in the weak field limit, we can expand the metric about Minkowski spacetime metric $\eta_{\mu \nu}$  by writing
\begin{equation}
    g_{\mu \nu} = \eta_{\mu \nu} + \frac{2}{m_p} h_{\mu \nu} \ ,
\end{equation}
where $h_{\mu\nu}$ is the dimensionful field associated to the graviton.\footnote{Note that in the GW literature, $h_{\mu\nu}$ is often the notation used to describe the dimensionless fluctuation of the metric, rather than the graviton field. Since we adopt a field theoretic approach, we use the corresponding convention of normalising with respect to the reduced Planck mass.}
Expanding the action in powers of $h_{\mu \nu}$, one can read off the well-known action for $h_{\mu \nu}$ (in transverse-traceless (TT) gauge)
\begin{equation}
    \mathcal{S} =  \int d^4 x \, \left[ -\frac{1}{2}(\partial_\mu h_{\rho \sigma})^2  -\frac{1}{m_p}  h_{\mu \nu} T^{\mu \nu} \right], 
\end{equation}
where $T^{\mu \nu}$ is the energy momentum tensor of the electromagnetic field, defined by
\begin{equation}\label{eq:EMTensor}
T^{\mu \nu} =  F^{\mu \alpha} F^{\nu}{}_{\alpha} - \frac{1}{4} \eta^{\mu \nu} F_{\alpha \beta} F^{\alpha \beta} .
\end{equation}

In Ref.~\cite{McDonald:2023ohd} it was shown how the coupling between axions and photons leads to a resonant conversion of axions into photons with a probability that can be read off simply by knowing the matrix element for axion to photon conversion, which can be immediately obtained from the interaction Lagrangian.
By generalising those results, one sees that the resonant conversion for a graviton into a photon can be written compactly as
\begin{equation}\label{eq:ConversionProb}
     P_{h \rightarrow \gamma} = 
     \frac{\pi \left| \mathcal{M}_{h \rightarrow \gamma } \right|^2}{E_\gamma \left| \textbf{k} \cdot \nabla E_\gamma \right|} \frac{U_E}{U} ,
\end{equation}
where $\left| \mathcal{M}_{h \rightarrow \gamma} \right|^2 $ is the squared matrix element for the conversion of gravitons into photons, $E_\gamma$ is the photon energy,  $\nabla E_\gamma$ is its spacial gradient, and $U_E$ and $U$ are the electric and total electromagnetic energy density in the photon mode. It is understood that all quantities are then evaluated ``on resonance" i.e. at the point where the photon and graviton dispersion relations become degenerate, with both satisfying $\omega = |\textbf{k}|$ where $\omega$ and $\textbf{k}$ are their frequency and 3-momentum, respectively. Full details can be found in Ref.~\cite{McDonald:2023ohd}.  We emphasise here that there is nothing intrinsically ``quantum" in our treatment of gravitational waves conversion, the above language is simply a useful method for computing conversion probabilities and fluxes. Indeed, the equivalence of the above approach and full solutions to classical wave equations was recently demonstrated in Ref.~\cite{Gines:2024ekm}.

Hence, to determine the resonant conversion probability for gravitons, one must simply compute the matrix element appearing in the numerator of Eq.~\eqref{eq:ConversionProb}.  To do this, we make further use of our TT gauge choice for the graviton polarisation tensors\footnote{We adopt a quantisation of the graviton field in terms of 4-momentum eigenstates with associated polarisation tensors, such that $h_{\mu\nu} \sim \sum_i\int d^3\vec{k}/((2\pi)^3\sqrt{2 \omega}) a_i (k)\,H_{\mu\nu}(k)e^{i k_\alpha x^\alpha} + \text{c.c.}$} $H_{\mu \nu}(k)$, which implies the conditions $k_{\mu} H^{\mu \nu }  =0 $ (transverse) and $H^\mu_{\, \, \mu} =0$ (traceless). We also expand $F_{\mu \nu}$ about a background field of the neutron star $F_{\rm NS}^{\mu \nu}$ by taking $F^{\mu \nu } \rightarrow F^{\mu \nu }_{\rm NS} + \partial^{[\mu } A^{\nu] }$ where $[\,]$ denotes antisymmetrisation and $A^\mu$ denotes the dynamical photon field that mixes with the graviton. Putting this together, the second term in  Eq.~\eqref{eq:EMTensor} does not contribute
when contracted with $h^{\mu \nu}$ via the traceless condition,
leaving
an effective interaction Lagrangian (using the transverse condition)
\begin{equation}
    \mathcal{L}_{\rm int} = \frac{2}{m_p} h_{ \mu \nu } \partial_{\alpha} A^{\mu}   F^{\nu \alpha}_{\rm NS} \ ,
\end{equation}
from which we can easily read off the squared matrix element for graviton to photon conversion as
\begin{equation}
    \left| \mathcal{M}^{+, \times}_{h \rightarrow \gamma } \right|^2 = \frac{1}{m_p^2}\left| 2 H^{+ , \times}_{\mu \nu} k_\alpha \epsilon^\mu F_{\rm NS}^{\nu \alpha} \right|^2,
\end{equation}
where $\epsilon$ is the (in-medium) polarisation 4-vector of the photon mode into which the graviton converts. Explicitly, working in temporal gauge for the photon in which $\epsilon^0 =0$, and assuming the presence of only a background magnetic field $B_k$ such that the non-vanishing components of $F_{\rm NS}$ are $F_{\rm NS}^{ij} =  \epsilon^{ijk} B^k$, 
we can write the full conversion probability as 
\begin{equation}
    P^{+, \times}_{h \rightarrow \gamma} = 
     \frac{4 \pi \left| \hat{\boldsymbol{\epsilon}} \cdot \textbf{H}^{+, \times} \cdot (\textbf{k}\times \textbf{B})\right|^2}{E_\gamma \left| \textbf{k} \cdot \nabla E_\gamma \right|} \frac{U_E}{U} \frac{1}{m_p^2},
\end{equation}
where $\hat{\boldsymbol{\epsilon}}$ is the electric field polarisation 3-vector.

The resonance occurs where the axion and photon 3-momenta become degenerate, which occurs when 
\begin{equation}\label{eq:ResCondition}
E_\gamma  = \left|\textbf{k}\right| \ .
\end{equation}
This condition is met on surfaces where the photon dispersion relation becomes null. To determine where these regions lie, we must first compute the dispersion relation for the photon. This is determined by the photon permittivity $\varepsilon$, which consists of two contributions 
\begin{equation}
    \varepsilon = \varepsilon_{\rm pl} + \varepsilon_{\rm vac }.
\end{equation}
The first, $\varepsilon_{\rm pl}$, is the standard permittivity of a classical magnetised plasma and $\varepsilon_{\rm vac}$ arises from quantum loop corrections to the photon self-energy in the presence of an external magnetic field. Explicitly we can choose coordinates in which $\textbf{k} = (0,0,k)$ and \cite{VasiliiSBeskin1986}
\begin{equation}
\varepsilon_{\rm pl} = R^{yz}_{\theta} \cdot \begin{pmatrix}
    \epsilon      & i g & 0  \\
    - i g & \epsilon & 0  \\
    0 & 0 & \eta
\end{pmatrix} \cdot R^{yz}_{- \theta} ,
\end{equation}
where the magnetic field is taken to be at an angle $\theta$ from the $z$-axis in the positive $y-z$ quadrant, and $R^{yz}_{\theta}$ is the rotation matrix by $\theta$ in the $y-z$-plane. The coefficients in the dielectric tensor are given by
\begin{equation}
\epsilon = 1 - \frac{\omega_p^2}{\omega^2 - \Omega_c^2},\quad g = \frac{\omega_p^2 \Omega_c}{\omega (\omega^2 - \Omega_c^2)}, \quad \eta = 1 - \frac{\omega_p^2}{\omega^2} 
\end{equation}
where $\omega_p = \sqrt{4 \pi \alpha n_e/m_e}$ and $\Omega_c = \sqrt{\alpha} B/m_e$ are the plasma frequency and cylotron frequency, respectively. The Euler-Heisenberg \cite{Heisenberg:1936nmg} contribution to the permittivity from the vacuum in the limit where $B < B_c \equiv m_e^2 /e$ is sub-critical is given by \cite{Adler:1971wn}
\begin{equation}
    \varepsilon_{\rm vac} = \mathbb{I}\left( 1 - \frac{8\alpha^2}{45 m_e^4} \left| \textbf{B} \right|^2\right) + \frac{28 \alpha^2}{45 m_e^3} \textbf{B} \otimes \textbf{B}\ .
\end{equation}
In addition, the vacuum birefringence introduces corrections to the magnetic permeability, which reads
\begin{equation}
    \mu_{ij}^{-1} = \mathbb{I}\left( 1 - \frac{8\alpha^2}{45 m_e^4} \left| \textbf{B} \right|^2\right) - \frac{16 \alpha^2}{45 m_e^3} \textbf{B} \otimes \textbf{B} \ .
\end{equation}
Substituting these expressions into Maxwell's equations and Fourier transforming, we find simple analytic expressions for the photon dispersion relation in two limits, $\Omega_c \ll \omega$ and $\Omega_c \gg \omega$. For the limit $\Omega_c \gg \omega_{\rm p}, \omega$, the plasma becomes strongly magnetised, and there are three modes \cite{PhysRevE.57.3399}: the magnetosonic-t, Langmuir-O (LO) and  Alfv{\'e}n modes, respectively. Only the LO mode is capable of propagating out of the plasma and escaping to vacuum at infinity. It has a refractive index-squared given by
\begin{equation}
    n^2_{\rm LO} =
            \frac{5 \left(\left(4 b^2+9\right) \omega ^2-9 \omega _p^2\right)}{\cos ^2(\theta ) \left(28 b^2 \omega ^2-45 \omega
   _p^2\right)+\left(45-8 b^2\right) \omega ^2} ,
\end{equation}
where $b = B/(m_e^2/\alpha)$. Meanwhile, for the limit $\omega \gg \Omega_c,\omega_{\rm p}$ we can send $\Omega_c /\omega$ to zero, and expand perturbatively in $\omega_{\rm p}/\omega$ and $b$. This gives two modes 
\begin{align}
n_\perp ^2 & = 1 - \frac{\omega_p^2}{\omega^2} + \frac{16}{45} b^2 \sin^2\theta , \nonumber \\
n_{\parallel}^2 & = 1 - \frac{\omega_p^2}{\omega^2} + \frac{28}{45} b^2 \sin^2 \theta, \label{eq:dispersionWeakB}
\end{align}
where the mode corresponding to  $n_\perp$ is polarised perpendicular to $\textbf{B}$ while $n_\parallel$ has both parallel \textit{and} perpendicular components relative to $\textbf{B}$. The resonance condition is met when Eq.~\eqref{eq:ResCondition} holds for the given photon mode. This corresponds to surfaces on which
\begin{equation}
    \omega_p^2 =  (28, \, 16 \sin^2 \theta, \, 28 \sin^2 \theta)
\frac{ \alpha^2 \omega^2 \left| \textbf{B}\right|^2 }{45 m_e^4},
\end{equation}
for the (${\rm LO}$, $\perp$, $\parallel$) modes, respectively. To finally obtain the conversion probability, we need to specify a basis for the polarisation vectors $H^{+,\times}$. 
We choose one in which the two graviton polarisations can be written as
\begin{equation}
    H^+  = \frac{1}{\sqrt{2}}
\begin{pmatrix}
	1 & \phantom{-}0 & 0 \\
	0 & -1 & 0 \\
	0 & \phantom{-}0 & 0
\end{pmatrix}, 
\qquad
 H^\times  = \frac{1}{\sqrt{2}}
\begin{pmatrix}
	0 & 1 & 0 \\
	1 & 0 & 0 \\
	0 & 0 & 0
\end{pmatrix}
.
\end{equation}

For the LO mode the conversion probabilities are thus given by
\begin{align}
    P^{\times}_{h \rightarrow {\rm LO}} &= \pi \frac{ \sin^2 \theta_B \left| \textbf{B} \right|^2 }{\left| \textbf{k} \cdot \nabla E_{\rm LO} \right|} \frac{\omega}{m_p^2}, \label{eq:PCross}\\
   P^{+}_{h \rightarrow \gamma}  &= 0.
\end{align}
with the gradient given by
\begin{equation}
    \nabla E_{\rm LO} = \frac{7 \omega\omega_p \sin ^2 \theta  }{7 \omega^2-7\omega_p^2 \cos ^2\theta +5\omega_p^2} \nabla \omega_p.
\end{equation}
Meanwhile, in the high-frequency regime, for our choice of polarisation basis, we find that $+$ converts exclusively to $\perp$ and $\parallel$ converts exclusively to $\times$, so that
\begin{align}
    &P^{+,\times}_{ h \rightarrow \gamma_{\perp,\parallel}} =   \pi \frac{ \sin^2 \theta_B \left| \textbf{B} \right|^2 }{\left| \textbf{k} \cdot \nabla E_{\perp,\parallel}\right|} \frac{\omega}{m_p^2}, \label{eq:PHighOmega}\\
   &P^{+}_{h \rightarrow \parallel}  = P^{\times}_{h \rightarrow \perp}= 0,
\end{align}
with similar expressions for the gradients which can be read off from the dispersion relations \eqref{eq:dispersionWeakB}. For all modes in question, $U_E/U = 1/2$.

We have now gathered together all the ingredients we need to compute the conversion probability. In the next section, we apply this to the magnetosphere of neutron stars. 

\section{Neutron Stars and Photon Flux}\label{sec:Neutorn Stars}
\begin{figure}[h!]
    \centering
    \includegraphics[scale=1.2]{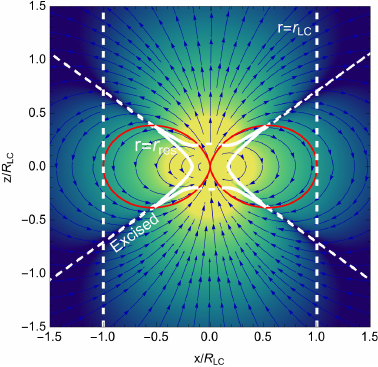}
    \caption{\textbf{Magnetosphere and Critical Surface.} The white solid line displays the critical surface for the LO mode corresponding to $\omega_p^2 = \frac{ 28 \alpha^2 \omega^2 \left| \textbf{B}\right|^2 }{45 m_e^4}$ in the strong magnetisation case $\Omega_c \gg \omega$ for an aligned-rotator ($\alpha = 0$) with $B_0= 10^{13}$G, $P=2\pi\,\text{s}$ and a frequency $\omega$ = keV. White dashed lines are excised from our analysis (see main text). The background density gives the Goldreich-Julian number density $|n_{GJ}|$ in arbitrary units and the red line displays the closed magnetic field line region. Dashed vertical lines correspond to the light cylinder at $R_{\rm LC} = 1/\Omega$.  }
    \label{fig:CriticalSurfacce}
\end{figure}

We begin by considering the canonical model for the magnetosphere of neutron stars, namely a Goldreich-Julian (GJ) \cite{Goldreich:1969sb} plasma distribution, where the number density of charge carriers is given by
\begin{align}
n_{\mathrm{GJ}}(\mathbf{r}) = \frac{2\, \boldsymbol{\Omega} \cdot \mathbf{B}}{e} \frac{1}{1 - \Omega^2 \,r^2\, \sin^2 \theta}\,,
\label{eq:nGJ}
\end{align}
where the magnetic field is given by a magnetic dipole rotating with angular frequency $\Omega = 2\pi/P$, inclinded at an angle $\alpha$ relative to the rotation axis:
\begin{align} \label{B-eq1}
    B_r & = B_0\left(\frac{R}{r}\right)^3(\cos \alpha \cos \theta+\sin \alpha \sin \theta \cos \psi), \\ \label{B-eq2}
    B_\theta  & =\frac{B_0}{2}\left(\frac{R}{r}\right)^3(\cos \alpha \sin \theta-\sin \alpha \cos \theta \cos \psi), \\ \label{B-eq3}
    B_\phi & =\frac{B_0}{2}\left(\frac{R}{r}\right)^3 \sin \alpha \sin \psi \ .
\end{align}
The quantity $\boldsymbol{\Omega} $ is the constant NS rotation vector, $B_0$ is the surface magnetic field strength, $\psi = \phi - \Omega t $ and $(r,\theta, \phi)$ are polar coordinates with the north pole given by the rotation axis. The plasma mass 
is then
\begin{equation}\label{eq:WP}
\omega^{\rm GJ}_{\rm p}(\vec{r}) = \sqrt{ \frac{4 \pi \, \alpha_{\rm EM} \, \left| n_{\rm GJ}(\vec{r}) \right|  }{m_{\rm e}} }.
\end{equation}
With this model, we can plot the critical surface (white solid line) in Fig~\ref{fig:CriticalSurfacce}. To give a conservative treatment of the magnetosphere, we have restricted our study of the critical surface to those parts which lie within the closed magnetic field lines (red line in Fig.~\ref{fig:CriticalSurfacce}) where the GJ model is expected to be a good approximation of the structure of the magnetosphere. The excised regions are shown as white dashed lines. We have also applied a cutoff condition to the integral in Eq.~\eqref{eq:FluxDensity} whenever the magnetic field strengths on the critical surface come within $10\%$ of the critical magnetic field strength, ensuring the perturbative nature of the Euler-Heisenberg expansion is respected.

Note that the diagonal lines in Fig.~\ref{fig:CriticalSurfacce} correspond to the so-called null surfaces in the GJ model on which $\omega^{\rm GJ}_{\rm p} \simeq 0$. In a realistic magnetosphere, these regions are expected to be filled with charges, making these sections of the critical surface inferred from the GJ model untrustworthy. 
To address this issue, for comparison, we also consider a simpler model in which $|n| = 2 \Omega B_0 (R/r)^3$, which captures the relevant scaling whilst avoiding complicated angular dependencies and spurious null surfaces present in the GJ model. We also use a further model (discussed more below) to capture charge densities which exceed GJ values, as might occur in magnetars. 

The total photon luminosity from gravitational wave conversion is given by \cite{McDonald:2023ohd}
\begin{equation}\label{eq:Luminosity}
    L  = \int d^3 \mathbf{k} \int d \Sigma_\mathbf{k} \cdot \mathbf{v}_p \, P_{h \rightarrow \gamma} \, \omega f_h,
\end{equation}
where $d\Sigma_\textbf{k}$ is the area elements on on the conversion surface associated to gravitons with momentum $\textbf{k}$.\footnote{Recall that owing to the anisotropic nature of the plasma, gravitons with different momenta convert on different surfaces, defining a foliation of resonant conversion surfaces. See Ref.~\cite{McDonald:2023shx}.} The phase velocity is given by $\textbf{v}_p = \textbf{k}/\omega$, $d^3\textbf{k}$ is the phase space measure for graviton 3-momenta and $f_h = f_h(\textbf{x},\textbf{k})$ is the graviton phase space density. 

\begin{figure}[ht]
    \centering
    \includegraphics[scale=0.85]{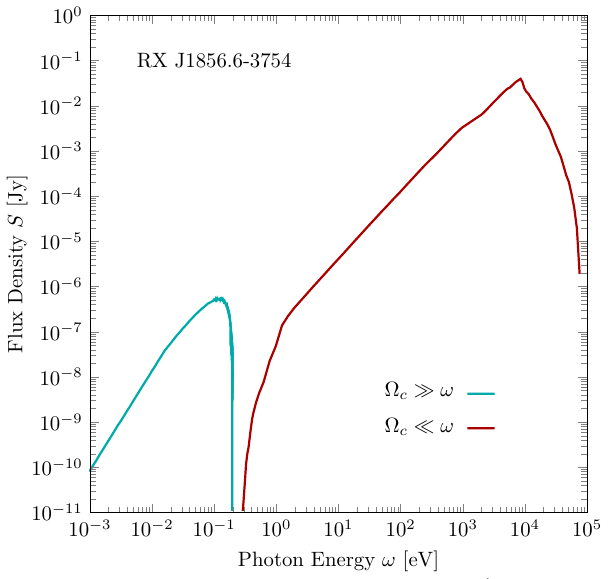}
    \caption{\textbf{Flux Density.} We display the average photon flux density Eq.~\eqref{eq:FluxDensity2} from gravitational waves as  function of frequency for a strain of $h_{c,\rm sto} = 10^{-20}$ for RX J1856.6-3754 which has $B_0 = 2.9\times 10^{13}$G, $P = 7.06\,\text{s}$ and $d = 123$pc. The cyan displays the flux corresponding to the probability \eqref{eq:PCross} whilst the red corresponds to the $\times \rightarrow \parallel$ processes of Eq.~\eqref{eq:PHighOmega}.  The middle cyan (red) asymptotes at $\omega \sim 10^{-1} {\rm eV}$   correspond to our cutting prescription on the value of $\omega$ set by $ \Omega_c > 10\, \omega$  ($10\, \Omega_c <  \omega$) in the low (high) frequency limits. The right-most red asymptote corresponds to demanding that the critical surface be inside the closed field line zone.  }
    \label{fig:FluxDensity}
\end{figure}

Next we write $d^3\textbf{k} = d\Omega_\textbf{k} d\omega \omega^2$, where $d\Omega_\textbf{k}$ is the solid angle for the graviton momentum in polar coordinates.  We must also remember to convert from angular to ordinary frequency, by using $\int d\omega = 2 \pi \int df$. For a source at a distance $d$ from Earth, this allows us to write down the photon flux density averaged over all emissions directions: 
\begin{equation}\label{eq:FluxDensity}
    S = \frac{1}{4 \pi \, d^2} \int d \Omega_{\mathbf{k}} \int d \boldsymbol{\Sigma}_{\mathbf{k}} \cdot \mathbf{v}_p 
    P_{h \rightarrow \gamma} \, 2\pi \omega^3 f_h
\end{equation}
Next we need to substitute the expression for the photon phase space distribution $f_h$ for the dimensionless strain of stochastic gravitational waves $h_{\rm c, sto}$. The energy density of gravitons is given by integrating $f_h$ over momentum space. Assuming an isotropic and homogeneous distribution of stochastic gravitational waves such that $f_h(\textbf{k},\textbf{x}) = f_h(\omega)$, we can write
\begin{equation}
    \rho_{\rm GW} = \int d^3 \textbf{k} \, \omega f_h = \int d \ln  f \, 4 \pi \omega^4 f_h,
\end{equation}
where again we used $d\omega = 2 \pi d f = \omega \, d \ln f  $. From this, we read off $d \rho_{\rm GW}/d\ln \omega = 4\pi \omega^4 f_h$ which by definition \cite{Aggarwal:2020olq}, is equal to $\rho_c \Omega_{\rm GW} = \pi f^2 h_{\rm c, sto}^2/(2G)  = \omega^2 h_{\rm c, sto}^2/(8 \pi G)  $. This enables us to identify $f_h =  h_{\rm c, sto}^2 /(32 \pi^2  G \omega^2) $. Inserting this expression for $f_h$ into Eq.~\eqref{eq:FluxDensity} we obtain an expression for the photon flux density in terms of the strain
\begin{equation}\label{eq:FluxDensity2}
    S = \frac{1}{ 4 \pi d^2} \int d \Omega_{\mathbf{k}} \int d \boldsymbol{\Sigma}_{\mathbf{k}} \cdot \mathbf{v}_p 
    P_{h \rightarrow \gamma}  \frac{\omega}{16 \pi G} h_{\rm c,sto}^2.
\end{equation}
The flux density \eqref{eq:FluxDensity2} is shown in Fig.~\ref{fig:FluxDensity} for a nearby isolated neutron star RX J1856.6-3754.

Note that at high $\omega$, the critical surface lies far from the star and as we increase $\omega$ further, the whole critical surface is pushed outside the closed magnetic field line region (red line Fig.~\ref{fig:CriticalSurfacce}). 
In the interest of being conservative, we excise contributions to the photon flux from such regions, so our signal vanishes by construction at high frequencies.
At low $\omega$, the toroidal contributions are pushed towards and inside the star, meaning that the surface becomes increasingly dominated by conical regions close to the so-called null surfaces where $n_{\rm GJ} \simeq 0$.  Great caution is therefore needed, since such null surfaces may not appear in realistic magnetosphere  models. However, we expect the conversion surface to partially track the low plasma density contours, with the main difference being that the precise morphology may differ from what is inferred from the GJ model. More detailed magnetosphere models should be used to address such uncertainties in future work. Nonetheless, our spherical plasma model (which does not suffer from such issues) also leads to strong constraints, suggesting that modifications to the canonical GJ model will not significantly affect the qualitative results, while only slightly affecting quantitative findings. 

\begin{figure*}
    \centering
    \includegraphics[width=\textwidth]{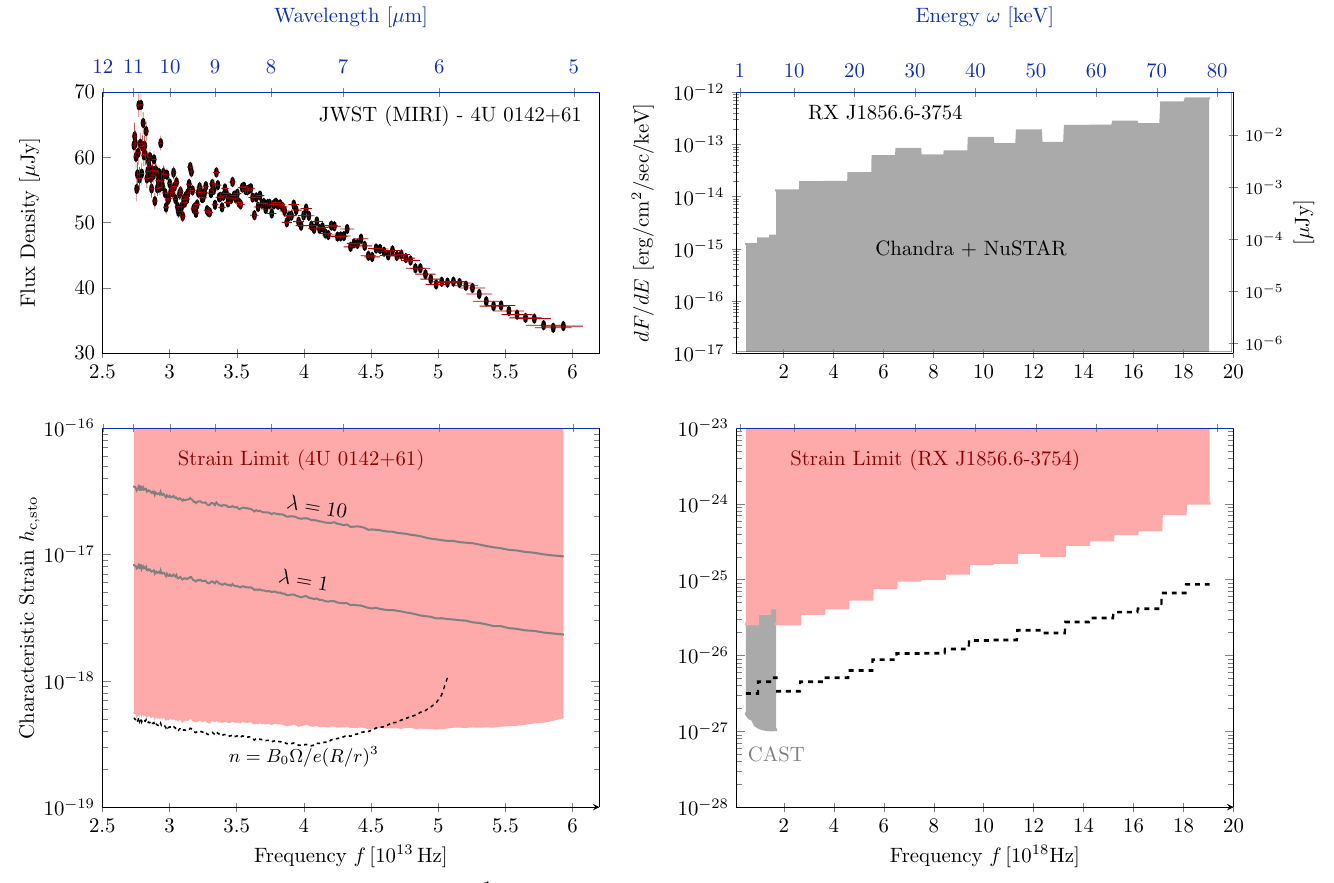}
    \caption{\textbf{Flux Measurements and Strain Limits. } \textit{Left Column.} In the top panel we show spectral measurements of 4U 0142+61 using JWST \cite{Hare:2024lcv} and corresponding constraints on the characteristic strain in the panel below. \textit{Right Column.} In the top panel we show flux limits from RX J1856.6-3754 in the energy ranges $2 - 8$keV (Chandra) \cite{Dessert:2019dos} and 3 - 79 keV using (NuStar), and in the bottom panel the corresponding constraints on the characteristic strain. Pink corresponds to a Goldreich Julian plasma profile \eqref{eq:nGJ}, whilst in both strain plots, black dashed lines indicate the same constraints with a simpler spherical profile corresponding to $|n| = \Omega B_0/e (R/r)^3$ in which angular dependencies (and hence GJ null surfaces), are removed. In the case of the magnetar 4U 0142+61, we follow \cite{McDonald:2023shx} (see references therein) and implement a Magnetar model with number density $n =  \lambda \frac{\psi}{e r} \sin^2\theta \, B_0 \, \left(R/r \right)^3$ with $\psi =0.2$ and $\lambda = 1, 10$, which can exceed the minimal GJ charge density. }
    \label{fig:total_flux}
\end{figure*}

\section{Constraints}\label{sec:Constraints}

The flux density from GW conversion in a neutron star magnetosphere (Eq.~\eqref{eq:FluxDensity2}) can be compared with observed fluxes to set constraints on the characteristic strain of the stochastic GW background. 

\begin{figure}[ht]
    \centering
    \includegraphics[scale=0.85]{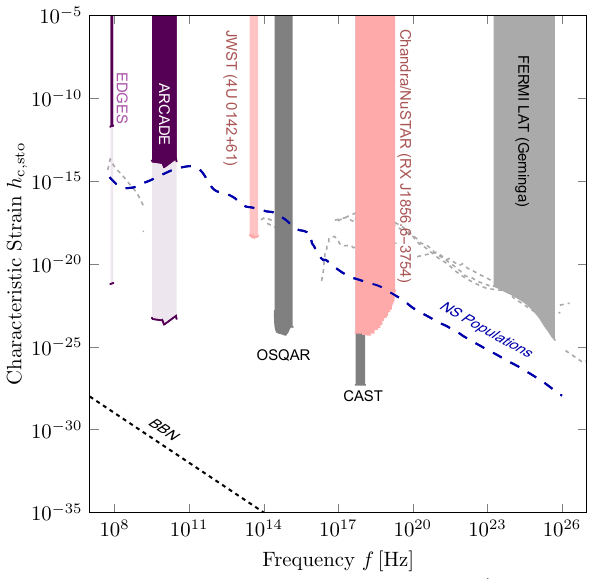}
    \caption{\textbf{Constraints on Characteristic Strain.} We show our strain limits (pink) from JWST, NuSTAR and Chandra alongside other constraints. We show laboratory constraints from OSQAR and CAST \cite{Ejlli:2019bqj} and other neutron star limits from Geminga \cite{Ito:2023fcr} for non-resonant conversion using FERMI-LAT data \cite{Fermi-LAT:2010mou}. We also show other neutron stars limits from Crab and Geminga \cite{Ito:2023fcr} as gray dashed lines (note there may not be continuous frequency coverage in these bands). The blue dashed line indicates limits from non-resonant conversion with neutron star populations, using the more conservative decaying magnetic field scenario \cite{Dandoy:2024oqg}, which again inherits gaps in its frequency coverage from the underlying data \cite{Hill:2018trh}. We also display constraints from the CMB \cite{Domcke:2020yzq} showing the optimistic (light purple) and pessimistic (dark purple) limits dependent on assumptions on the background magnetic fields. }
    \label{fig:Constraint}
\end{figure}

In the X-ray band, observations of RX J1856.6-3754 by the Chandra~\cite{10.1117/12.671760,Dessert:2019dos} and NuSTAR~\cite{2013ApJ...770..103H,NuStarHEASARC} telescopes can be used to set constraints in the range $ 5\times 10^{17}\,\text{Hz} \lesssim f \lesssim 2\times 10^{19}\,\text{Hz}$. For the Chandra data, a thorough analysis searching for analogous signals from axion conversion in the neutron star magnetosphere has already been performed, with results presented in Ref.~\cite{Dessert:2019dos}. We recast the results of the measured flux in the energy range $2 - 8\,\text{keV}$ as a constraint on stochastic GWs, as shown in the right panel of Fig.~\ref{fig:total_flux}. For NuSTAR data, we use the observed flux in the energy range $3-79\,\text{keV}$ from 47ks of observation time reported in the HEASARC catalog~\cite{NuStarHEASARC}. In order to obtain an approximate flux density, we use the effective area as a function of photon energy reported in Ref.~\cite{2013ApJ...770..103H}. The resultant flux density is shown in the upper-right panel of Fig.~\ref{fig:total_flux}. The corresopnding constraint on $h_c$ is shown in the lower-right panel of the same figure, and in the context of the wider frequency range of GWs in Fig.~\ref{fig:Constraint}. We caution that our analysis of NuSTAR X-ray data is not as thorough as that performed in Ref.~\cite{Dessert:2019dos} for the Chandra data. However, since there are possibly significant systematic errors affecting the signal, including astrophysical, the present level of data analysis seems appropriate for setting constraints, and we defer a more thorough analysis of NuSTAR data to future work.

In the IR band, we make use of JWST spectral measurements of the magnetar 4U 0142$+$61 from Ref.~\cite{Hare:2024lcv}.\footnote{We thank the authors for providing us with their data.} These constraints must be interpreted with caution, as the magnetosphere of a magnetar may differ from the GJ model we use to compute the expected signal flux density. We attempt to capture this via a model considered in Ref.~\cite{McDonald:2023shx} which attempts to quantify the increase in charge density above the minimal value set by the GJ model with a number density given by $n =  \lambda \frac{\psi}{e r} \sin^2\theta \, B_0 \, \left(R/r \right)^3$ where $\psi$ and $\lambda$ are constants.  With this in mind, it is nonethtless instructive to examine the size of strain sensitivity with a view to further JWST observations of other (non-magnetar) neutron star spectra, or more systematic treatments of the magnetosphere. We display results in the left column of FIG.~\ref{fig:total_flux}.

Constraints using \textit{non-resonant} conversion have previously been explored in Ref.~\cite{Ito:2023fcr} using observations of the Crab and Geminga pulsars. In Fig.~\ref{fig:Constraint} we display results for Geminga data from FERMI-LAT found in Ref.~\cite{Fermi-LAT:2010mou} wherein bin widths are clearly stated and the data covers a continuous 
range $0.1 \lesssim \omega \lesssim 34 {\rm GeV}$. Other Neutron star constraints \cite{Ito:2023fcr, Dandoy:2024oqg} are displayed as dashed lines to emphasise that they are derived from data which may not cover a continuous frequency range. These constraints use, e.g., earlier data from the Compton Gamma Ray Observatory of Geminga and Crab (see Ref.~\cite{1996ASPC..105..307T}) for which it is unclear whether data covers a continuous frequency range, nor are the bin widths clear. 
Similar comments apply to FERMI-LAT observations of Crab \cite{Meyer:2010tta}, so these are also shown as dashed lines.
We also caution that the magnetosphere treatment and graviton-photon mixing in those works was far simpler that what is presented here. 

Non-resonant conversion limits from neutron star populations, reported in Ref.~\cite{Dandoy:2024oqg}, are also shown, though we caution that they formally have incomplete frequency coverage owing to gaps in the underlying spectral measurements \cite{Hill:2018trh}. We display results for the more conservative populations scenario where the magnetic field strength decays in time. We also illustrate limits/projections from various laboratory experiments \cite{Ejlli:2019bqj}. We see that resonant conversion of GWs into photons in single neutron stars leads to stronger limits than the non-resonant conversion from the full galactic neutron star population study. Therefore, it is plausible that a population study, looking for resonant conversion, would result in further improved sensitivity. We leave this study to future work.

Ultimately, while resonant conversion in neutron star magnetospheres offers impressive sensitivities to stochastic sources, it should be cautioned that the origin of a signal with the corresponding amplitude would have to be from the late universe. Indeed, constraints on the number of relativistic degrees of freedom during the era of Big Bang Nucleosynthesis (BBN) restrict $\Omega_{\rm GW} \lesssim 10^{-6}$, corresponding to $h_c \lesssim 10^{-30} \times (\text{GHz}/f)$~\cite{Yeh:2022heq}. The result is that at the frequencies relevant to the individual neutron stars we have considered, the BBN bound constrains early universe GWs to have characteristic strains at least 14 orders of magnitude smaller than what can be probed by either 4U 0142+61 or RX J1856.6-3754.

Late universe stochastic GWs could arise from, e.g., unresolved PBH mergers, although the associated spectrum is expected to have a peak amplitude lower than what can be probed by our study~\cite{Franciolini:2022htd}. Ultimately, since the study of UHFGWs is still in its infancy, the full range of sources, especially from late universe processes, is not known.

\section{Summary}\label{sec:conclusions}

In this work we have outlined a new mechanism for resonantly converting high-frequency gravitational waves into photons in the magnetospheres of neutron stars. This exploits the inverse Gertsenshtein effect and strong magnetic fields in addition to a resonance which occurs in regions where the photon has a null dispersion relation. We have calculated the production rate at different frequencies and described the morphology of the magnetosphere surfaces on which resonant conversion can take place within the Goldreich-Julian model \cite{Goldreich:1969sb}, as well as a spherical plasma and magnetar-like \cite{McDonald:2023shx} model.  We have seen that at low frequencies, the relevant mode into which gravitons convert is the LO mode, and at higher frequencies, there are two photon production modes, with resonant conversion occurring on a foliation of surfaces with the shape of each member of the foliation depending on the angle between the gravitational wave 3-momentum and the magnetic field. Clearly it will be interesting to carry out a more detailed study of magnetosphere models to understand how these affect the flux density from resonant gravitational wave conversion.

We obtained limits on the characteristic gravitational waves strain $h_{c,\rm sto}$ by first computing the expected photon flux from gravitational waves converting into photons in the magnetosphere of the neutron star. Formally, there is a systematic error on constraints from the observing angle $\theta$ which is the angle subtending the rotation axis of the star and the line of sight to the observer. Such uncertainties can be quantified through ray-tracing \cite{McDonald:2023shx}, though we leave such a detailed analysis for future work. We used the prediction for the photon flux to set limits on the size of the characteristic strain $h_c$ by comparing to observations of neutron stars in the X-ray and IR. We obtained competitive constraints on the stochastic strain $h_c$ which exceed both existing limits both from individual stars \cite{Ito:2023fcr} and populations \cite{Dandoy:2024oqg} by many orders of magnitude. 

In future work it would be beneficial to carry out a more exhaustive study of all archival X-ray and IR data across a wider range of neutron stars, as well as considering telescope data in other frequency ranges. This leaves the door open to a stronger and wider range of constraints on high frequency gravitational waves.

\section*{Acknowledgments}

We thank Francesca Chadha-Day, Juraj Claric,  Virgile Dandoy, Valerie Domcke, Camilo Garcia-Cely, Bettina Posselt and Sam Witte for useful conversations. We are grateful to  Asuka Ito and Virgile Dandoy for sharing data. We also thank Jeremy Hare, George Pavlov, Bettina Posselt, Oleg Kargaltsev, Tea Temim and Steven Chen for supplying JWST data on 4U 0142+6 used from their work \cite{Hare:2024lcv}. Aldo Ejlli has also kindly provided us with CAST and OSQAR limits from his work \cite{Ejlli:2019bqj}. This work has benefited from discussions held at the workshop ``Ultra-high frequency graviational waves: where to next?" which took place in CERN,  funded by the CERN-Korea Theory Collaboration and by the UKRI/EPSRC Stephen Hawking fellowship, grant reference EP/T017279/1. JM thanks CERN for hospitality and financial support form a Manchester University Research Collaboration fund.  The work of SARE was supported by SNF Ambizione grant PZ00P2\_193322, \textit{New frontiers from sub-eV to super-TeV}.

\bibliography{references}

\begin{thebibliography}{83}%
\makeatletter
\providecommand \@ifxundefined [1]{%
 \@ifx{#1\undefined}
}%
\providecommand \@ifnum [1]{%
 \ifnum #1\expandafter \@firstoftwo
 \else \expandafter \@secondoftwo
 \fi
}%
\providecommand \@ifx [1]{%
 \ifx #1\expandafter \@firstoftwo
 \else \expandafter \@secondoftwo
 \fi
}%
\providecommand \natexlab [1]{#1}%
\providecommand \enquote  [1]{``#1''}%
\providecommand \bibnamefont  [1]{#1}%
\providecommand \bibfnamefont [1]{#1}%
\providecommand \citenamefont [1]{#1}%
\providecommand \href@noop [0]{\@secondoftwo}%
\providecommand \href [0]{\begingroup \@sanitize@url \@href}%
\providecommand \@href[1]{\@@startlink{#1}\@@href}%
\providecommand \@@href[1]{\endgroup#1\@@endlink}%
\providecommand \@sanitize@url [0]{\catcode `\\12\catcode `\$12\catcode `\&12\catcode `\#12\catcode `\^12\catcode `\_12\catcode `\%12\relax}%
\providecommand \@@startlink[1]{}%
\providecommand \@@endlink[0]{}%
\providecommand \url  [0]{\begingroup\@sanitize@url \@url }%
\providecommand \@url [1]{\endgroup\@href {#1}{\urlprefix }}%
\providecommand \urlprefix  [0]{URL }%
\providecommand \Eprint [0]{\href }%
\providecommand \doibase [0]{http://dx.doi.org/}%
\providecommand \selectlanguage [0]{\@gobble}%
\providecommand \bibinfo  [0]{\@secondoftwo}%
\providecommand \bibfield  [0]{\@secondoftwo}%
\providecommand \translation [1]{[#1]}%
\providecommand \BibitemOpen [0]{}%
\providecommand \bibitemStop [0]{}%
\providecommand \bibitemNoStop [0]{.\EOS\space}%
\providecommand \EOS [0]{\spacefactor3000\relax}%
\providecommand \BibitemShut  [1]{\csname bibitem#1\endcsname}%
\let\auto@bib@innerbib\@empty
\bibitem [{\citenamefont {Abbott}\ \emph {et~al.}(2016{\natexlab{a}})\citenamefont {Abbott} \emph {et~al.}}]{LIGOScientific:2016aoc}%
  \BibitemOpen
  \bibfield  {author} {\bibinfo {author} {\bibfnamefont {B.~P.}\ \bibnamefont {Abbott}} \emph {et~al.} (\bibinfo {collaboration} {LIGO Scientific, Virgo}),\ }\href {\doibase 10.1103/PhysRevLett.116.061102} {\bibfield  {journal} {\bibinfo  {journal} {Phys. Rev. Lett.}\ }\textbf {\bibinfo {volume} {116}},\ \bibinfo {pages} {061102} (\bibinfo {year} {2016}{\natexlab{a}})},\ \Eprint {http://arxiv.org/abs/1602.03837} {arXiv:1602.03837 [gr-qc]} \BibitemShut {NoStop}%
\bibitem [{\citenamefont {Abbott}\ \emph {et~al.}(2016{\natexlab{b}})\citenamefont {Abbott} \emph {et~al.}}]{LIGOScientific:2016emj}%
  \BibitemOpen
  \bibfield  {author} {\bibinfo {author} {\bibfnamefont {B.~P.}\ \bibnamefont {Abbott}} \emph {et~al.} (\bibinfo {collaboration} {LIGO Scientific, Virgo}),\ }\href {\doibase 10.1103/PhysRevLett.116.131103} {\bibfield  {journal} {\bibinfo  {journal} {Phys. Rev. Lett.}\ }\textbf {\bibinfo {volume} {116}},\ \bibinfo {pages} {131103} (\bibinfo {year} {2016}{\natexlab{b}})},\ \Eprint {http://arxiv.org/abs/1602.03838} {arXiv:1602.03838 [gr-qc]} \BibitemShut {NoStop}%
\bibitem [{\citenamefont {Acernese}\ \emph {et~al.}(2015)\citenamefont {Acernese} \emph {et~al.}}]{VIRGO:2014yos}%
  \BibitemOpen
  \bibfield  {author} {\bibinfo {author} {\bibfnamefont {F.}~\bibnamefont {Acernese}} \emph {et~al.} (\bibinfo {collaboration} {VIRGO}),\ }\href {\doibase 10.1088/0264-9381/32/2/024001} {\bibfield  {journal} {\bibinfo  {journal} {Class. Quant. Grav.}\ }\textbf {\bibinfo {volume} {32}},\ \bibinfo {pages} {024001} (\bibinfo {year} {2015})},\ \Eprint {http://arxiv.org/abs/1408.3978} {arXiv:1408.3978 [gr-qc]} \BibitemShut {NoStop}%
\bibitem [{\citenamefont {Aso}\ \emph {et~al.}(2013)\citenamefont {Aso}, \citenamefont {Michimura}, \citenamefont {Somiya}, \citenamefont {Ando}, \citenamefont {Miyakawa}, \citenamefont {Sekiguchi}, \citenamefont {Tatsumi},\ and\ \citenamefont {Yamamoto}}]{Aso:2013eba}%
  \BibitemOpen
  \bibfield  {author} {\bibinfo {author} {\bibfnamefont {Y.}~\bibnamefont {Aso}}, \bibinfo {author} {\bibfnamefont {Y.}~\bibnamefont {Michimura}}, \bibinfo {author} {\bibfnamefont {K.}~\bibnamefont {Somiya}}, \bibinfo {author} {\bibfnamefont {M.}~\bibnamefont {Ando}}, \bibinfo {author} {\bibfnamefont {O.}~\bibnamefont {Miyakawa}}, \bibinfo {author} {\bibfnamefont {T.}~\bibnamefont {Sekiguchi}}, \bibinfo {author} {\bibfnamefont {D.}~\bibnamefont {Tatsumi}}, \ and\ \bibinfo {author} {\bibfnamefont {H.}~\bibnamefont {Yamamoto}} (\bibinfo {collaboration} {KAGRA}),\ }\href {\doibase 10.1103/PhysRevD.88.043007} {\bibfield  {journal} {\bibinfo  {journal} {Phys. Rev. D}\ }\textbf {\bibinfo {volume} {88}},\ \bibinfo {pages} {043007} (\bibinfo {year} {2013})},\ \Eprint {http://arxiv.org/abs/1306.6747} {arXiv:1306.6747 [gr-qc]} \BibitemShut {NoStop}%
\bibitem [{\citenamefont {Arzoumanian}\ \emph {et~al.}(2020)\citenamefont {Arzoumanian} \emph {et~al.}}]{NANOGrav:2020bcs}%
  \BibitemOpen
  \bibfield  {author} {\bibinfo {author} {\bibfnamefont {Z.}~\bibnamefont {Arzoumanian}} \emph {et~al.} (\bibinfo {collaboration} {NANOGrav}),\ }\href {\doibase 10.3847/2041-8213/abd401} {\bibfield  {journal} {\bibinfo  {journal} {Astrophys. J. Lett.}\ }\textbf {\bibinfo {volume} {905}},\ \bibinfo {pages} {L34} (\bibinfo {year} {2020})},\ \Eprint {http://arxiv.org/abs/2009.04496} {arXiv:2009.04496 [astro-ph.HE]} \BibitemShut {NoStop}%
\bibitem [{\citenamefont {Goncharov}\ \emph {et~al.}(2021)\citenamefont {Goncharov} \emph {et~al.}}]{Goncharov:2021oub}%
  \BibitemOpen
  \bibfield  {author} {\bibinfo {author} {\bibfnamefont {B.}~\bibnamefont {Goncharov}} \emph {et~al.},\ }\href {\doibase 10.3847/2041-8213/ac17f4} {\bibfield  {journal} {\bibinfo  {journal} {Astrophys. J. Lett.}\ }\textbf {\bibinfo {volume} {917}},\ \bibinfo {pages} {L19} (\bibinfo {year} {2021})},\ \Eprint {http://arxiv.org/abs/2107.12112} {arXiv:2107.12112 [astro-ph.HE]} \BibitemShut {NoStop}%
\bibitem [{\citenamefont {Chen}\ \emph {et~al.}(2021)\citenamefont {Chen} \emph {et~al.}}]{EPTA:2021crs}%
  \BibitemOpen
  \bibfield  {author} {\bibinfo {author} {\bibfnamefont {S.}~\bibnamefont {Chen}} \emph {et~al.} (\bibinfo {collaboration} {EPTA}),\ }\href {\doibase 10.1093/mnras/stab2833} {\bibfield  {journal} {\bibinfo  {journal} {Mon. Not. Roy. Astron. Soc.}\ }\textbf {\bibinfo {volume} {508}},\ \bibinfo {pages} {4970} (\bibinfo {year} {2021})},\ \Eprint {http://arxiv.org/abs/2110.13184} {arXiv:2110.13184 [astro-ph.HE]} \BibitemShut {NoStop}%
\bibitem [{\citenamefont {Tarafdar}\ \emph {et~al.}(2022)\citenamefont {Tarafdar} \emph {et~al.}}]{Tarafdar:2022toa}%
  \BibitemOpen
  \bibfield  {author} {\bibinfo {author} {\bibfnamefont {P.}~\bibnamefont {Tarafdar}} \emph {et~al.},\ }\href {\doibase 10.1017/pasa.2022.46} {\bibfield  {journal} {\bibinfo  {journal} {Publ. Astron. Soc. Austral.}\ }\textbf {\bibinfo {volume} {39}},\ \bibinfo {pages} {e053} (\bibinfo {year} {2022})},\ \Eprint {http://arxiv.org/abs/2206.09289} {arXiv:2206.09289 [astro-ph.IM]} \BibitemShut {NoStop}%
\bibitem [{\citenamefont {Amaro-Seoane}\ \emph {et~al.}(2017)\citenamefont {Amaro-Seoane} \emph {et~al.}}]{LISA:2017pwj}%
  \BibitemOpen
  \bibfield  {author} {\bibinfo {author} {\bibfnamefont {P.}~\bibnamefont {Amaro-Seoane}} \emph {et~al.} (\bibinfo {collaboration} {LISA}),\ }\href@noop {} {\  (\bibinfo {year} {2017})},\ \Eprint {http://arxiv.org/abs/1702.00786} {arXiv:1702.00786 [astro-ph.IM]} \BibitemShut {NoStop}%
\bibitem [{\citenamefont {Badurina}\ \emph {et~al.}(2020)\citenamefont {Badurina} \emph {et~al.}}]{Badurina:2019hst}%
  \BibitemOpen
  \bibfield  {author} {\bibinfo {author} {\bibfnamefont {L.}~\bibnamefont {Badurina}} \emph {et~al.},\ }\href {\doibase 10.1088/1475-7516/2020/05/011} {\bibfield  {journal} {\bibinfo  {journal} {JCAP}\ }\textbf {\bibinfo {volume} {05}},\ \bibinfo {pages} {011} (\bibinfo {year} {2020})},\ \Eprint {http://arxiv.org/abs/1911.11755} {arXiv:1911.11755 [astro-ph.CO]} \BibitemShut {NoStop}%
\bibitem [{\citenamefont {Abe}\ \emph {et~al.}(2021)\citenamefont {Abe} \emph {et~al.}}]{MAGIS-100:2021etm}%
  \BibitemOpen
  \bibfield  {author} {\bibinfo {author} {\bibfnamefont {M.}~\bibnamefont {Abe}} \emph {et~al.} (\bibinfo {collaboration} {MAGIS-100}),\ }\href {\doibase 10.1088/2058-9565/abf719} {\bibfield  {journal} {\bibinfo  {journal} {Quantum Sci. Technol.}\ }\textbf {\bibinfo {volume} {6}},\ \bibinfo {pages} {044003} (\bibinfo {year} {2021})},\ \Eprint {http://arxiv.org/abs/2104.02835} {arXiv:2104.02835 [physics.atom-ph]} \BibitemShut {NoStop}%
\bibitem [{\citenamefont {Rubin}\ and\ \citenamefont {Ford}(1970)}]{Rubin:1970zza}%
  \BibitemOpen
  \bibfield  {author} {\bibinfo {author} {\bibfnamefont {V.~C.}\ \bibnamefont {Rubin}}\ and\ \bibinfo {author} {\bibfnamefont {W.~K.}\ \bibnamefont {Ford}, \bibfnamefont {Jr.}},\ }\href {\doibase 10.1086/150317} {\bibfield  {journal} {\bibinfo  {journal} {Astrophys. J.}\ }\textbf {\bibinfo {volume} {159}},\ \bibinfo {pages} {379} (\bibinfo {year} {1970})}\BibitemShut {NoStop}%
\bibitem [{\citenamefont {Perlmutter}\ \emph {et~al.}(1999)\citenamefont {Perlmutter} \emph {et~al.}}]{SupernovaCosmologyProject:1998vns}%
  \BibitemOpen
  \bibfield  {author} {\bibinfo {author} {\bibfnamefont {S.}~\bibnamefont {Perlmutter}} \emph {et~al.} (\bibinfo {collaboration} {Supernova Cosmology Project}),\ }\href {\doibase 10.1086/307221} {\bibfield  {journal} {\bibinfo  {journal} {Astrophys. J.}\ }\textbf {\bibinfo {volume} {517}},\ \bibinfo {pages} {565} (\bibinfo {year} {1999})},\ \Eprint {http://arxiv.org/abs/astro-ph/9812133} {arXiv:astro-ph/9812133} \BibitemShut {NoStop}%
\bibitem [{\citenamefont {Riess}\ \emph {et~al.}(1998)\citenamefont {Riess} \emph {et~al.}}]{SupernovaSearchTeam:1998fmf}%
  \BibitemOpen
  \bibfield  {author} {\bibinfo {author} {\bibfnamefont {A.~G.}\ \bibnamefont {Riess}} \emph {et~al.} (\bibinfo {collaboration} {Supernova Search Team}),\ }\href {\doibase 10.1086/300499} {\bibfield  {journal} {\bibinfo  {journal} {Astron. J.}\ }\textbf {\bibinfo {volume} {116}},\ \bibinfo {pages} {1009} (\bibinfo {year} {1998})},\ \Eprint {http://arxiv.org/abs/astro-ph/9805201} {arXiv:astro-ph/9805201} \BibitemShut {NoStop}%
\bibitem [{\citenamefont {Penzias}\ and\ \citenamefont {Wilson}(1965)}]{Penzias:1965wn}%
  \BibitemOpen
  \bibfield  {author} {\bibinfo {author} {\bibfnamefont {A.~A.}\ \bibnamefont {Penzias}}\ and\ \bibinfo {author} {\bibfnamefont {R.~W.}\ \bibnamefont {Wilson}},\ }\href {\doibase 10.1086/148307} {\bibfield  {journal} {\bibinfo  {journal} {Astrophys. J.}\ }\textbf {\bibinfo {volume} {142}},\ \bibinfo {pages} {419} (\bibinfo {year} {1965})}\BibitemShut {NoStop}%
\bibitem [{\citenamefont {Aggarwal}\ \emph {et~al.}(2021)\citenamefont {Aggarwal} \emph {et~al.}}]{Aggarwal:2020olq}%
  \BibitemOpen
  \bibfield  {author} {\bibinfo {author} {\bibfnamefont {N.}~\bibnamefont {Aggarwal}} \emph {et~al.},\ }\href {\doibase 10.1007/s41114-021-00032-5} {\bibfield  {journal} {\bibinfo  {journal} {Living Rev. Rel.}\ }\textbf {\bibinfo {volume} {24}},\ \bibinfo {pages} {4} (\bibinfo {year} {2021})},\ \Eprint {http://arxiv.org/abs/2011.12414} {arXiv:2011.12414 [gr-qc]} \BibitemShut {NoStop}%
\bibitem [{\citenamefont {Servant}\ and\ \citenamefont {Simakachorn}(2024)}]{Servant:2023tua}%
  \BibitemOpen
  \bibfield  {author} {\bibinfo {author} {\bibfnamefont {G.}~\bibnamefont {Servant}}\ and\ \bibinfo {author} {\bibfnamefont {P.}~\bibnamefont {Simakachorn}},\ }\href {\doibase 10.1103/PhysRevD.109.103538} {\bibfield  {journal} {\bibinfo  {journal} {Phys. Rev. D}\ }\textbf {\bibinfo {volume} {109}},\ \bibinfo {pages} {103538} (\bibinfo {year} {2024})},\ \Eprint {http://arxiv.org/abs/2312.09281} {arXiv:2312.09281 [hep-ph]} \BibitemShut {NoStop}%
\bibitem [{\citenamefont {Hindmarsh}\ \emph {et~al.}(2017)\citenamefont {Hindmarsh}, \citenamefont {Huber}, \citenamefont {Rummukainen},\ and\ \citenamefont {Weir}}]{Hindmarsh:2017gnf}%
  \BibitemOpen
  \bibfield  {author} {\bibinfo {author} {\bibfnamefont {M.}~\bibnamefont {Hindmarsh}}, \bibinfo {author} {\bibfnamefont {S.~J.}\ \bibnamefont {Huber}}, \bibinfo {author} {\bibfnamefont {K.}~\bibnamefont {Rummukainen}}, \ and\ \bibinfo {author} {\bibfnamefont {D.~J.}\ \bibnamefont {Weir}},\ }\href {\doibase 10.1103/PhysRevD.96.103520} {\bibfield  {journal} {\bibinfo  {journal} {Phys. Rev. D}\ }\textbf {\bibinfo {volume} {96}},\ \bibinfo {pages} {103520} (\bibinfo {year} {2017})},\ \bibinfo {note} {[Erratum: Phys.Rev.D 101, 089902 (2020)]},\ \Eprint {http://arxiv.org/abs/1704.05871} {arXiv:1704.05871 [astro-ph.CO]} \BibitemShut {NoStop}%
\bibitem [{\citenamefont {Ghiglieri}\ and\ \citenamefont {Laine}(2015)}]{Ghiglieri:2015nfa}%
  \BibitemOpen
  \bibfield  {author} {\bibinfo {author} {\bibfnamefont {J.}~\bibnamefont {Ghiglieri}}\ and\ \bibinfo {author} {\bibfnamefont {M.}~\bibnamefont {Laine}},\ }\href {\doibase 10.1088/1475-7516/2015/07/022} {\bibfield  {journal} {\bibinfo  {journal} {JCAP}\ }\textbf {\bibinfo {volume} {07}},\ \bibinfo {pages} {022} (\bibinfo {year} {2015})},\ \Eprint {http://arxiv.org/abs/1504.02569} {arXiv:1504.02569 [hep-ph]} \BibitemShut {NoStop}%
\bibitem [{\citenamefont {Ringwald}\ \emph {et~al.}(2021)\citenamefont {Ringwald}, \citenamefont {Sch\"utte-Engel},\ and\ \citenamefont {Tamarit}}]{Ringwald:2020ist}%
  \BibitemOpen
  \bibfield  {author} {\bibinfo {author} {\bibfnamefont {A.}~\bibnamefont {Ringwald}}, \bibinfo {author} {\bibfnamefont {J.}~\bibnamefont {Sch\"utte-Engel}}, \ and\ \bibinfo {author} {\bibfnamefont {C.}~\bibnamefont {Tamarit}},\ }\href {\doibase 10.1088/1475-7516/2021/03/054} {\bibfield  {journal} {\bibinfo  {journal} {JCAP}\ }\textbf {\bibinfo {volume} {03}},\ \bibinfo {pages} {054} (\bibinfo {year} {2021})},\ \Eprint {http://arxiv.org/abs/2011.04731} {arXiv:2011.04731 [hep-ph]} \BibitemShut {NoStop}%
\bibitem [{\citenamefont {Hawking}(1971)}]{Hawking:1971ei}%
  \BibitemOpen
  \bibfield  {author} {\bibinfo {author} {\bibfnamefont {S.}~\bibnamefont {Hawking}},\ }\href {\doibase 10.1093/mnras/152.1.75} {\bibfield  {journal} {\bibinfo  {journal} {Mon. Not. Roy. Astron. Soc.}\ }\textbf {\bibinfo {volume} {152}},\ \bibinfo {pages} {75} (\bibinfo {year} {1971})}\BibitemShut {NoStop}%
\bibitem [{\citenamefont {Khlopov}(2010)}]{Khlopov:2008qy}%
  \BibitemOpen
  \bibfield  {author} {\bibinfo {author} {\bibfnamefont {M.~Y.}\ \bibnamefont {Khlopov}},\ }\href {\doibase 10.1088/1674-4527/10/6/001} {\bibfield  {journal} {\bibinfo  {journal} {Res. Astron. Astrophys.}\ }\textbf {\bibinfo {volume} {10}},\ \bibinfo {pages} {495} (\bibinfo {year} {2010})},\ \Eprint {http://arxiv.org/abs/0801.0116} {arXiv:0801.0116 [astro-ph]} \BibitemShut {NoStop}%
\bibitem [{\citenamefont {Carr}\ \emph {et~al.}(2021)\citenamefont {Carr}, \citenamefont {Kohri}, \citenamefont {Sendouda},\ and\ \citenamefont {Yokoyama}}]{Carr:2020gox}%
  \BibitemOpen
  \bibfield  {author} {\bibinfo {author} {\bibfnamefont {B.}~\bibnamefont {Carr}}, \bibinfo {author} {\bibfnamefont {K.}~\bibnamefont {Kohri}}, \bibinfo {author} {\bibfnamefont {Y.}~\bibnamefont {Sendouda}}, \ and\ \bibinfo {author} {\bibfnamefont {J.}~\bibnamefont {Yokoyama}},\ }\href {\doibase 10.1088/1361-6633/ac1e31} {\bibfield  {journal} {\bibinfo  {journal} {Rept. Prog. Phys.}\ }\textbf {\bibinfo {volume} {84}},\ \bibinfo {pages} {116902} (\bibinfo {year} {2021})},\ \Eprint {http://arxiv.org/abs/2002.12778} {arXiv:2002.12778 [astro-ph.CO]} \BibitemShut {NoStop}%
\bibitem [{\citenamefont {Carr}\ and\ \citenamefont {Kuhnel}(2022)}]{Carr:2021bzv}%
  \BibitemOpen
  \bibfield  {author} {\bibinfo {author} {\bibfnamefont {B.}~\bibnamefont {Carr}}\ and\ \bibinfo {author} {\bibfnamefont {F.}~\bibnamefont {Kuhnel}},\ }\href {\doibase 10.21468/SciPostPhysLectNotes.48} {\bibfield  {journal} {\bibinfo  {journal} {SciPost Phys. Lect. Notes}\ }\textbf {\bibinfo {volume} {48}},\ \bibinfo {pages} {1} (\bibinfo {year} {2022})},\ \Eprint {http://arxiv.org/abs/2110.02821} {arXiv:2110.02821 [astro-ph.CO]} \BibitemShut {NoStop}%
\bibitem [{\citenamefont {Franciolini}\ \emph {et~al.}(2022)\citenamefont {Franciolini}, \citenamefont {Maharana},\ and\ \citenamefont {Muia}}]{Franciolini:2022htd}%
  \BibitemOpen
  \bibfield  {author} {\bibinfo {author} {\bibfnamefont {G.}~\bibnamefont {Franciolini}}, \bibinfo {author} {\bibfnamefont {A.}~\bibnamefont {Maharana}}, \ and\ \bibinfo {author} {\bibfnamefont {F.}~\bibnamefont {Muia}},\ }\href {\doibase 10.1103/PhysRevD.106.103520} {\bibfield  {journal} {\bibinfo  {journal} {Phys. Rev. D}\ }\textbf {\bibinfo {volume} {106}},\ \bibinfo {pages} {103520} (\bibinfo {year} {2022})},\ \Eprint {http://arxiv.org/abs/2205.02153} {arXiv:2205.02153 [astro-ph.CO]} \BibitemShut {NoStop}%
\bibitem [{\citenamefont {Ternov}\ \emph {et~al.}(1978)\citenamefont {Ternov}, \citenamefont {Khalilov}, \citenamefont {Chizhov},\ and\ \citenamefont {Gaina}}]{Ternov:1978gq}%
  \BibitemOpen
  \bibfield  {author} {\bibinfo {author} {\bibfnamefont {I.~M.}\ \bibnamefont {Ternov}}, \bibinfo {author} {\bibfnamefont {V.~R.}\ \bibnamefont {Khalilov}}, \bibinfo {author} {\bibfnamefont {G.~A.}\ \bibnamefont {Chizhov}}, \ and\ \bibinfo {author} {\bibfnamefont {A.~B.}\ \bibnamefont {Gaina}},\ }\href {\doibase 10.1007/BF00894575} {\bibfield  {journal} {\bibinfo  {journal} {Sov. Phys. J.}\ }\textbf {\bibinfo {volume} {21}},\ \bibinfo {pages} {1200} (\bibinfo {year} {1978})}\BibitemShut {NoStop}%
\bibitem [{\citenamefont {Zouros}\ and\ \citenamefont {Eardley}(1979)}]{Zouros:1979iw}%
  \BibitemOpen
  \bibfield  {author} {\bibinfo {author} {\bibfnamefont {T.~J.~M.}\ \bibnamefont {Zouros}}\ and\ \bibinfo {author} {\bibfnamefont {D.~M.}\ \bibnamefont {Eardley}},\ }\href {\doibase 10.1016/0003-4916(79)90237-9} {\bibfield  {journal} {\bibinfo  {journal} {Annals Phys.}\ }\textbf {\bibinfo {volume} {118}},\ \bibinfo {pages} {139} (\bibinfo {year} {1979})}\BibitemShut {NoStop}%
\bibitem [{\citenamefont {Detweiler}(1980)}]{Detweiler:1980uk}%
  \BibitemOpen
  \bibfield  {author} {\bibinfo {author} {\bibfnamefont {S.~L.}\ \bibnamefont {Detweiler}},\ }\href {\doibase 10.1103/PhysRevD.22.2323} {\bibfield  {journal} {\bibinfo  {journal} {Phys. Rev. D}\ }\textbf {\bibinfo {volume} {22}},\ \bibinfo {pages} {2323} (\bibinfo {year} {1980})}\BibitemShut {NoStop}%
\bibitem [{\citenamefont {Arvanitaki}\ \emph {et~al.}(2010)\citenamefont {Arvanitaki}, \citenamefont {Dimopoulos}, \citenamefont {Dubovsky}, \citenamefont {Kaloper},\ and\ \citenamefont {March-Russell}}]{Arvanitaki:2009fg}%
  \BibitemOpen
  \bibfield  {author} {\bibinfo {author} {\bibfnamefont {A.}~\bibnamefont {Arvanitaki}}, \bibinfo {author} {\bibfnamefont {S.}~\bibnamefont {Dimopoulos}}, \bibinfo {author} {\bibfnamefont {S.}~\bibnamefont {Dubovsky}}, \bibinfo {author} {\bibfnamefont {N.}~\bibnamefont {Kaloper}}, \ and\ \bibinfo {author} {\bibfnamefont {J.}~\bibnamefont {March-Russell}},\ }\href {\doibase 10.1103/PhysRevD.81.123530} {\bibfield  {journal} {\bibinfo  {journal} {Phys. Rev. D}\ }\textbf {\bibinfo {volume} {81}},\ \bibinfo {pages} {123530} (\bibinfo {year} {2010})},\ \Eprint {http://arxiv.org/abs/0905.4720} {arXiv:0905.4720 [hep-th]} \BibitemShut {NoStop}%
\bibitem [{\citenamefont {Arvanitaki}\ and\ \citenamefont {Dubovsky}(2011)}]{Arvanitaki:2010sy}%
  \BibitemOpen
  \bibfield  {author} {\bibinfo {author} {\bibfnamefont {A.}~\bibnamefont {Arvanitaki}}\ and\ \bibinfo {author} {\bibfnamefont {S.}~\bibnamefont {Dubovsky}},\ }\href {\doibase 10.1103/PhysRevD.83.044026} {\bibfield  {journal} {\bibinfo  {journal} {Phys. Rev. D}\ }\textbf {\bibinfo {volume} {83}},\ \bibinfo {pages} {044026} (\bibinfo {year} {2011})},\ \Eprint {http://arxiv.org/abs/1004.3558} {arXiv:1004.3558 [hep-th]} \BibitemShut {NoStop}%
\bibitem [{\citenamefont {Yoshino}\ and\ \citenamefont {Kodama}(2014)}]{Yoshino:2013ofa}%
  \BibitemOpen
  \bibfield  {author} {\bibinfo {author} {\bibfnamefont {H.}~\bibnamefont {Yoshino}}\ and\ \bibinfo {author} {\bibfnamefont {H.}~\bibnamefont {Kodama}},\ }\href {\doibase 10.1093/ptep/ptu029} {\bibfield  {journal} {\bibinfo  {journal} {PTEP}\ }\textbf {\bibinfo {volume} {2014}},\ \bibinfo {pages} {043E02} (\bibinfo {year} {2014})},\ \Eprint {http://arxiv.org/abs/1312.2326} {arXiv:1312.2326 [gr-qc]} \BibitemShut {NoStop}%
\bibitem [{\citenamefont {Brito}\ \emph {et~al.}(2015{\natexlab{a}})\citenamefont {Brito}, \citenamefont {Cardoso},\ and\ \citenamefont {Pani}}]{Brito:2014wla}%
  \BibitemOpen
  \bibfield  {author} {\bibinfo {author} {\bibfnamefont {R.}~\bibnamefont {Brito}}, \bibinfo {author} {\bibfnamefont {V.}~\bibnamefont {Cardoso}}, \ and\ \bibinfo {author} {\bibfnamefont {P.}~\bibnamefont {Pani}},\ }\href {\doibase 10.1088/0264-9381/32/13/134001} {\bibfield  {journal} {\bibinfo  {journal} {Class. Quant. Grav.}\ }\textbf {\bibinfo {volume} {32}},\ \bibinfo {pages} {134001} (\bibinfo {year} {2015}{\natexlab{a}})},\ \Eprint {http://arxiv.org/abs/1411.0686} {arXiv:1411.0686 [gr-qc]} \BibitemShut {NoStop}%
\bibitem [{\citenamefont {Arvanitaki}\ \emph {et~al.}(2015)\citenamefont {Arvanitaki}, \citenamefont {Baryakhtar},\ and\ \citenamefont {Huang}}]{Arvanitaki:2014wva}%
  \BibitemOpen
  \bibfield  {author} {\bibinfo {author} {\bibfnamefont {A.}~\bibnamefont {Arvanitaki}}, \bibinfo {author} {\bibfnamefont {M.}~\bibnamefont {Baryakhtar}}, \ and\ \bibinfo {author} {\bibfnamefont {X.}~\bibnamefont {Huang}},\ }\href {\doibase 10.1103/PhysRevD.91.084011} {\bibfield  {journal} {\bibinfo  {journal} {Phys. Rev. D}\ }\textbf {\bibinfo {volume} {91}},\ \bibinfo {pages} {084011} (\bibinfo {year} {2015})},\ \Eprint {http://arxiv.org/abs/1411.2263} {arXiv:1411.2263 [hep-ph]} \BibitemShut {NoStop}%
\bibitem [{\citenamefont {Brito}\ \emph {et~al.}(2015{\natexlab{b}})\citenamefont {Brito}, \citenamefont {Cardoso},\ and\ \citenamefont {Pani}}]{Brito:2015oca}%
  \BibitemOpen
  \bibfield  {author} {\bibinfo {author} {\bibfnamefont {R.}~\bibnamefont {Brito}}, \bibinfo {author} {\bibfnamefont {V.}~\bibnamefont {Cardoso}}, \ and\ \bibinfo {author} {\bibfnamefont {P.}~\bibnamefont {Pani}},\ }\href {\doibase 10.1007/978-3-319-19000-6} {\bibfield  {journal} {\bibinfo  {journal} {Lect. Notes Phys.}\ }\textbf {\bibinfo {volume} {906}},\ \bibinfo {pages} {pp.1} (\bibinfo {year} {2015}{\natexlab{b}})},\ \Eprint {http://arxiv.org/abs/1501.06570} {arXiv:1501.06570 [gr-qc]} \BibitemShut {NoStop}%
\bibitem [{\citenamefont {Zhu}\ \emph {et~al.}(2020)\citenamefont {Zhu}, \citenamefont {Baryakhtar}, \citenamefont {Papa}, \citenamefont {Tsuna}, \citenamefont {Kawanaka},\ and\ \citenamefont {Eggenstein}}]{Zhu:2020tht}%
  \BibitemOpen
  \bibfield  {author} {\bibinfo {author} {\bibfnamefont {S.~J.}\ \bibnamefont {Zhu}}, \bibinfo {author} {\bibfnamefont {M.}~\bibnamefont {Baryakhtar}}, \bibinfo {author} {\bibfnamefont {M.~A.}\ \bibnamefont {Papa}}, \bibinfo {author} {\bibfnamefont {D.}~\bibnamefont {Tsuna}}, \bibinfo {author} {\bibfnamefont {N.}~\bibnamefont {Kawanaka}}, \ and\ \bibinfo {author} {\bibfnamefont {H.-B.}\ \bibnamefont {Eggenstein}},\ }\href {\doibase 10.1103/PhysRevD.102.063020} {\bibfield  {journal} {\bibinfo  {journal} {Phys. Rev. D}\ }\textbf {\bibinfo {volume} {102}},\ \bibinfo {pages} {063020} (\bibinfo {year} {2020})},\ \Eprint {http://arxiv.org/abs/2003.03359} {arXiv:2003.03359 [gr-qc]} \BibitemShut {NoStop}%
\bibitem [{\citenamefont {Prabhu}(2021)}]{Prabhu:2021zve}%
  \BibitemOpen
  \bibfield  {author} {\bibinfo {author} {\bibfnamefont {A.}~\bibnamefont {Prabhu}},\ }\href {\doibase 10.1103/PhysRevD.104.055038} {\bibfield  {journal} {\bibinfo  {journal} {Phys. Rev. D}\ }\textbf {\bibinfo {volume} {104}},\ \bibinfo {pages} {055038} (\bibinfo {year} {2021})},\ \Eprint {http://arxiv.org/abs/2104.14569} {arXiv:2104.14569 [hep-ph]} \BibitemShut {NoStop}%
\bibitem [{\citenamefont {Noordhuis}\ \emph {et~al.}(2023)\citenamefont {Noordhuis}, \citenamefont {Prabhu}, \citenamefont {Witte}, \citenamefont {Chen}, \citenamefont {Cruz},\ and\ \citenamefont {Weniger}}]{Noordhuis:2022ljw}%
  \BibitemOpen
  \bibfield  {author} {\bibinfo {author} {\bibfnamefont {D.}~\bibnamefont {Noordhuis}}, \bibinfo {author} {\bibfnamefont {A.}~\bibnamefont {Prabhu}}, \bibinfo {author} {\bibfnamefont {S.~J.}\ \bibnamefont {Witte}}, \bibinfo {author} {\bibfnamefont {A.~Y.}\ \bibnamefont {Chen}}, \bibinfo {author} {\bibfnamefont {F.}~\bibnamefont {Cruz}}, \ and\ \bibinfo {author} {\bibfnamefont {C.}~\bibnamefont {Weniger}},\ }\href {\doibase 10.1103/PhysRevLett.131.111004} {\bibfield  {journal} {\bibinfo  {journal} {Phys. Rev. Lett.}\ }\textbf {\bibinfo {volume} {131}},\ \bibinfo {pages} {111004} (\bibinfo {year} {2023})},\ \Eprint {http://arxiv.org/abs/2209.09917} {arXiv:2209.09917 [hep-ph]} \BibitemShut {NoStop}%
\bibitem [{\citenamefont {Witte}\ \emph {et~al.}(2024)\citenamefont {Witte}, \citenamefont {Noordhuis}, \citenamefont {Prabhu},\ and\ \citenamefont {Weniger}}]{Witte:2024akb}%
  \BibitemOpen
  \bibfield  {author} {\bibinfo {author} {\bibfnamefont {S.}~\bibnamefont {Witte}}, \bibinfo {author} {\bibfnamefont {D.}~\bibnamefont {Noordhuis}}, \bibinfo {author} {\bibfnamefont {A.}~\bibnamefont {Prabhu}}, \ and\ \bibinfo {author} {\bibfnamefont {C.}~\bibnamefont {Weniger}},\ }\href {\doibase 10.22323/1.454.0022} {\bibfield  {journal} {\bibinfo  {journal} {PoS}\ }\textbf {\bibinfo {volume} {COSMICWISPers}},\ \bibinfo {pages} {022} (\bibinfo {year} {2024})}\BibitemShut {NoStop}%
\bibitem [{\citenamefont {Casalderrey-Solana}\ \emph {et~al.}(2022)\citenamefont {Casalderrey-Solana}, \citenamefont {Mateos},\ and\ \citenamefont {Sanchez-Garitaonandia}}]{Casalderrey-Solana:2022rrn}%
  \BibitemOpen
  \bibfield  {author} {\bibinfo {author} {\bibfnamefont {J.}~\bibnamefont {Casalderrey-Solana}}, \bibinfo {author} {\bibfnamefont {D.}~\bibnamefont {Mateos}}, \ and\ \bibinfo {author} {\bibfnamefont {M.}~\bibnamefont {Sanchez-Garitaonandia}},\ }\href@noop {} {\  (\bibinfo {year} {2022})},\ \Eprint {http://arxiv.org/abs/2210.03171} {arXiv:2210.03171 [hep-th]} \BibitemShut {NoStop}%
\bibitem [{\citenamefont {Berlin}\ \emph {et~al.}(2022)\citenamefont {Berlin}, \citenamefont {Blas}, \citenamefont {Tito~D'Agnolo}, \citenamefont {Ellis}, \citenamefont {Harnik}, \citenamefont {Kahn},\ and\ \citenamefont {Sch\"utte-Engel}}]{Berlin:2021txa}%
  \BibitemOpen
  \bibfield  {author} {\bibinfo {author} {\bibfnamefont {A.}~\bibnamefont {Berlin}}, \bibinfo {author} {\bibfnamefont {D.}~\bibnamefont {Blas}}, \bibinfo {author} {\bibfnamefont {R.}~\bibnamefont {Tito~D'Agnolo}}, \bibinfo {author} {\bibfnamefont {S.~A.~R.}\ \bibnamefont {Ellis}}, \bibinfo {author} {\bibfnamefont {R.}~\bibnamefont {Harnik}}, \bibinfo {author} {\bibfnamefont {Y.}~\bibnamefont {Kahn}}, \ and\ \bibinfo {author} {\bibfnamefont {J.}~\bibnamefont {Sch\"utte-Engel}},\ }\href {\doibase 10.1103/PhysRevD.105.116011} {\bibfield  {journal} {\bibinfo  {journal} {Phys. Rev. D}\ }\textbf {\bibinfo {volume} {105}},\ \bibinfo {pages} {116011} (\bibinfo {year} {2022})},\ \Eprint {http://arxiv.org/abs/2112.11465} {arXiv:2112.11465 [hep-ph]} \BibitemShut {NoStop}%
\bibitem [{\citenamefont {Domcke}\ \emph {et~al.}(2022)\citenamefont {Domcke}, \citenamefont {Garcia-Cely},\ and\ \citenamefont {Rodd}}]{Domcke:2022rgu}%
  \BibitemOpen
  \bibfield  {author} {\bibinfo {author} {\bibfnamefont {V.}~\bibnamefont {Domcke}}, \bibinfo {author} {\bibfnamefont {C.}~\bibnamefont {Garcia-Cely}}, \ and\ \bibinfo {author} {\bibfnamefont {N.~L.}\ \bibnamefont {Rodd}},\ }\href {\doibase 10.1103/PhysRevLett.129.041101} {\bibfield  {journal} {\bibinfo  {journal} {Phys. Rev. Lett.}\ }\textbf {\bibinfo {volume} {129}},\ \bibinfo {pages} {041101} (\bibinfo {year} {2022})},\ \Eprint {http://arxiv.org/abs/2202.00695} {arXiv:2202.00695 [hep-ph]} \BibitemShut {NoStop}%
\bibitem [{\citenamefont {Arvanitaki}\ and\ \citenamefont {Geraci}(2013)}]{Arvanitaki:2012cn}%
  \BibitemOpen
  \bibfield  {author} {\bibinfo {author} {\bibfnamefont {A.}~\bibnamefont {Arvanitaki}}\ and\ \bibinfo {author} {\bibfnamefont {A.~A.}\ \bibnamefont {Geraci}},\ }\href {\doibase 10.1103/PhysRevLett.110.071105} {\bibfield  {journal} {\bibinfo  {journal} {Phys. Rev. Lett.}\ }\textbf {\bibinfo {volume} {110}},\ \bibinfo {pages} {071105} (\bibinfo {year} {2013})},\ \Eprint {http://arxiv.org/abs/1207.5320} {arXiv:1207.5320 [gr-qc]} \BibitemShut {NoStop}%
\bibitem [{\citenamefont {Aggarwal}\ \emph {et~al.}(2022)\citenamefont {Aggarwal}, \citenamefont {Winstone}, \citenamefont {Teo}, \citenamefont {Baryakhtar}, \citenamefont {Larson}, \citenamefont {Kalogera},\ and\ \citenamefont {Geraci}}]{Aggarwal:2020umq}%
  \BibitemOpen
  \bibfield  {author} {\bibinfo {author} {\bibfnamefont {N.}~\bibnamefont {Aggarwal}}, \bibinfo {author} {\bibfnamefont {G.~P.}\ \bibnamefont {Winstone}}, \bibinfo {author} {\bibfnamefont {M.}~\bibnamefont {Teo}}, \bibinfo {author} {\bibfnamefont {M.}~\bibnamefont {Baryakhtar}}, \bibinfo {author} {\bibfnamefont {S.~L.}\ \bibnamefont {Larson}}, \bibinfo {author} {\bibfnamefont {V.}~\bibnamefont {Kalogera}}, \ and\ \bibinfo {author} {\bibfnamefont {A.~A.}\ \bibnamefont {Geraci}},\ }\href {\doibase 10.1103/PhysRevLett.128.111101} {\bibfield  {journal} {\bibinfo  {journal} {Phys. Rev. Lett.}\ }\textbf {\bibinfo {volume} {128}},\ \bibinfo {pages} {111101} (\bibinfo {year} {2022})},\ \Eprint {http://arxiv.org/abs/2010.13157} {arXiv:2010.13157 [gr-qc]} \BibitemShut {NoStop}%
\bibitem [{\citenamefont {Goryachev}\ and\ \citenamefont {Tobar}(2014)}]{Goryachev:2014yra}%
  \BibitemOpen
  \bibfield  {author} {\bibinfo {author} {\bibfnamefont {M.}~\bibnamefont {Goryachev}}\ and\ \bibinfo {author} {\bibfnamefont {M.~E.}\ \bibnamefont {Tobar}},\ }\href {\doibase 10.1103/PhysRevD.90.102005} {\bibfield  {journal} {\bibinfo  {journal} {Phys. Rev. D}\ }\textbf {\bibinfo {volume} {90}},\ \bibinfo {pages} {102005} (\bibinfo {year} {2014})},\ \bibinfo {note} {[Erratum: Phys.Rev.D 108, 129901 (2023)]},\ \Eprint {http://arxiv.org/abs/1410.2334} {arXiv:1410.2334 [gr-qc]} \BibitemShut {NoStop}%
\bibitem [{\citenamefont {Bringmann}\ \emph {et~al.}(2023)\citenamefont {Bringmann}, \citenamefont {Domcke}, \citenamefont {Fuchs},\ and\ \citenamefont {Kopp}}]{Bringmann:2023gba}%
  \BibitemOpen
  \bibfield  {author} {\bibinfo {author} {\bibfnamefont {T.}~\bibnamefont {Bringmann}}, \bibinfo {author} {\bibfnamefont {V.}~\bibnamefont {Domcke}}, \bibinfo {author} {\bibfnamefont {E.}~\bibnamefont {Fuchs}}, \ and\ \bibinfo {author} {\bibfnamefont {J.}~\bibnamefont {Kopp}},\ }\href {\doibase 10.1103/PhysRevD.108.L061303} {\bibfield  {journal} {\bibinfo  {journal} {Phys. Rev. D}\ }\textbf {\bibinfo {volume} {108}},\ \bibinfo {pages} {L061303} (\bibinfo {year} {2023})},\ \Eprint {http://arxiv.org/abs/2304.10579} {arXiv:2304.10579 [hep-ph]} \BibitemShut {NoStop}%
\bibitem [{\citenamefont {Berlin}\ \emph {et~al.}(2023)\citenamefont {Berlin}, \citenamefont {Blas}, \citenamefont {Tito~D'Agnolo}, \citenamefont {Ellis}, \citenamefont {Harnik}, \citenamefont {Kahn}, \citenamefont {Sch\"utte-Engel},\ and\ \citenamefont {Wentzel}}]{Berlin:2023grv}%
  \BibitemOpen
  \bibfield  {author} {\bibinfo {author} {\bibfnamefont {A.}~\bibnamefont {Berlin}}, \bibinfo {author} {\bibfnamefont {D.}~\bibnamefont {Blas}}, \bibinfo {author} {\bibfnamefont {R.}~\bibnamefont {Tito~D'Agnolo}}, \bibinfo {author} {\bibfnamefont {S.~A.~R.}\ \bibnamefont {Ellis}}, \bibinfo {author} {\bibfnamefont {R.}~\bibnamefont {Harnik}}, \bibinfo {author} {\bibfnamefont {Y.}~\bibnamefont {Kahn}}, \bibinfo {author} {\bibfnamefont {J.}~\bibnamefont {Sch\"utte-Engel}}, \ and\ \bibinfo {author} {\bibfnamefont {M.}~\bibnamefont {Wentzel}},\ }\href {\doibase 10.1103/PhysRevD.108.084058} {\bibfield  {journal} {\bibinfo  {journal} {Phys. Rev. D}\ }\textbf {\bibinfo {volume} {108}},\ \bibinfo {pages} {084058} (\bibinfo {year} {2023})},\ \Eprint {http://arxiv.org/abs/2303.01518} {arXiv:2303.01518 [hep-ph]} \BibitemShut {NoStop}%
\bibitem [{\citenamefont {Baryakhtar}\ \emph {et~al.}(2022)\citenamefont {Baryakhtar} \emph {et~al.}}]{Baryakhtar:2022hbu}%
  \BibitemOpen
  \bibfield  {author} {\bibinfo {author} {\bibfnamefont {M.}~\bibnamefont {Baryakhtar}} \emph {et~al.},\ }in\ \href@noop {} {\emph {\bibinfo {booktitle} {{Snowmass 2021}}}}\ (\bibinfo {year} {2022})\ \Eprint {http://arxiv.org/abs/2203.07984} {arXiv:2203.07984 [hep-ph]} \BibitemShut {NoStop}%
\bibitem [{\citenamefont {Pshirkov}\ and\ \citenamefont {Popov}(2009)}]{Pshirkov:2007st}%
  \BibitemOpen
  \bibfield  {author} {\bibinfo {author} {\bibfnamefont {M.~S.}\ \bibnamefont {Pshirkov}}\ and\ \bibinfo {author} {\bibfnamefont {S.~B.}\ \bibnamefont {Popov}},\ }\href {\doibase 10.1134/S1063776109030030} {\bibfield  {journal} {\bibinfo  {journal} {J. Exp. Theor. Phys.}\ }\textbf {\bibinfo {volume} {108}},\ \bibinfo {pages} {384} (\bibinfo {year} {2009})},\ \Eprint {http://arxiv.org/abs/0711.1264} {arXiv:0711.1264 [astro-ph]} \BibitemShut {NoStop}%
\bibitem [{\citenamefont {Hook}\ \emph {et~al.}(2018)\citenamefont {Hook}, \citenamefont {Kahn}, \citenamefont {Safdi},\ and\ \citenamefont {Sun}}]{Hook:2018iia}%
  \BibitemOpen
  \bibfield  {author} {\bibinfo {author} {\bibfnamefont {A.}~\bibnamefont {Hook}}, \bibinfo {author} {\bibfnamefont {Y.}~\bibnamefont {Kahn}}, \bibinfo {author} {\bibfnamefont {B.~R.}\ \bibnamefont {Safdi}}, \ and\ \bibinfo {author} {\bibfnamefont {Z.}~\bibnamefont {Sun}},\ }\href {\doibase 10.1103/PhysRevLett.121.241102} {\bibfield  {journal} {\bibinfo  {journal} {Phys. Rev. Lett.}\ }\textbf {\bibinfo {volume} {121}},\ \bibinfo {pages} {241102} (\bibinfo {year} {2018})},\ \Eprint {http://arxiv.org/abs/1804.03145} {arXiv:1804.03145 [hep-ph]} \BibitemShut {NoStop}%
\bibitem [{\citenamefont {Huang}\ \emph {et~al.}(2018)\citenamefont {Huang}, \citenamefont {Kadota}, \citenamefont {Sekiguchi},\ and\ \citenamefont {Tashiro}}]{Huang:2018lxq}%
  \BibitemOpen
  \bibfield  {author} {\bibinfo {author} {\bibfnamefont {F.~P.}\ \bibnamefont {Huang}}, \bibinfo {author} {\bibfnamefont {K.}~\bibnamefont {Kadota}}, \bibinfo {author} {\bibfnamefont {T.}~\bibnamefont {Sekiguchi}}, \ and\ \bibinfo {author} {\bibfnamefont {H.}~\bibnamefont {Tashiro}},\ }\href {\doibase 10.1103/PhysRevD.97.123001} {\bibfield  {journal} {\bibinfo  {journal} {Phys. Rev. D}\ }\textbf {\bibinfo {volume} {97}},\ \bibinfo {pages} {123001} (\bibinfo {year} {2018})},\ \Eprint {http://arxiv.org/abs/1803.08230} {arXiv:1803.08230 [hep-ph]} \BibitemShut {NoStop}%
\bibitem [{\citenamefont {Battye}\ \emph {et~al.}(2020)\citenamefont {Battye}, \citenamefont {Garbrecht}, \citenamefont {McDonald}, \citenamefont {Pace},\ and\ \citenamefont {Srinivasan}}]{Battye:2019aco}%
  \BibitemOpen
  \bibfield  {author} {\bibinfo {author} {\bibfnamefont {R.~A.}\ \bibnamefont {Battye}}, \bibinfo {author} {\bibfnamefont {B.}~\bibnamefont {Garbrecht}}, \bibinfo {author} {\bibfnamefont {J.~I.}\ \bibnamefont {McDonald}}, \bibinfo {author} {\bibfnamefont {F.}~\bibnamefont {Pace}}, \ and\ \bibinfo {author} {\bibfnamefont {S.}~\bibnamefont {Srinivasan}},\ }\href {\doibase 10.1103/PhysRevD.102.023504} {\bibfield  {journal} {\bibinfo  {journal} {Phys. Rev. D}\ }\textbf {\bibinfo {volume} {102}},\ \bibinfo {pages} {023504} (\bibinfo {year} {2020})},\ \Eprint {http://arxiv.org/abs/1910.11907} {arXiv:1910.11907 [astro-ph.CO]} \BibitemShut {NoStop}%
\bibitem [{\citenamefont {Darling}(2020{\natexlab{a}})}]{Darling:2020uyo}%
  \BibitemOpen
  \bibfield  {author} {\bibinfo {author} {\bibfnamefont {J.}~\bibnamefont {Darling}},\ }\href {\doibase 10.3847/2041-8213/abb23f} {\bibfield  {journal} {\bibinfo  {journal} {Astrophys. J. Lett.}\ }\textbf {\bibinfo {volume} {900}},\ \bibinfo {pages} {L28} (\bibinfo {year} {2020}{\natexlab{a}})},\ \Eprint {http://arxiv.org/abs/2008.11188} {arXiv:2008.11188 [astro-ph.CO]} \BibitemShut {NoStop}%
\bibitem [{\citenamefont {Battye}\ \emph {et~al.}(2022)\citenamefont {Battye}, \citenamefont {Darling}, \citenamefont {McDonald},\ and\ \citenamefont {Srinivasan}}]{Battye:2021yue}%
  \BibitemOpen
  \bibfield  {author} {\bibinfo {author} {\bibfnamefont {R.~A.}\ \bibnamefont {Battye}}, \bibinfo {author} {\bibfnamefont {J.}~\bibnamefont {Darling}}, \bibinfo {author} {\bibfnamefont {J.~I.}\ \bibnamefont {McDonald}}, \ and\ \bibinfo {author} {\bibfnamefont {S.}~\bibnamefont {Srinivasan}},\ }\href {\doibase 10.1103/PhysRevD.105.L021305} {\bibfield  {journal} {\bibinfo  {journal} {Phys. Rev. D}\ }\textbf {\bibinfo {volume} {105}},\ \bibinfo {pages} {L021305} (\bibinfo {year} {2022})},\ \Eprint {http://arxiv.org/abs/2107.01225} {arXiv:2107.01225 [astro-ph.CO]} \BibitemShut {NoStop}%
\bibitem [{\citenamefont {Darling}(2020{\natexlab{b}})}]{Darling:2020plz}%
  \BibitemOpen
  \bibfield  {author} {\bibinfo {author} {\bibfnamefont {J.}~\bibnamefont {Darling}},\ }\href {\doibase 10.1103/PhysRevLett.125.121103} {\bibfield  {journal} {\bibinfo  {journal} {Phys. Rev. Lett.}\ }\textbf {\bibinfo {volume} {125}},\ \bibinfo {pages} {121103} (\bibinfo {year} {2020}{\natexlab{b}})},\ \Eprint {http://arxiv.org/abs/2008.01877} {arXiv:2008.01877 [astro-ph.CO]} \BibitemShut {NoStop}%
\bibitem [{\citenamefont {Foster}\ \emph {et~al.}(2022)\citenamefont {Foster}, \citenamefont {Witte}, \citenamefont {Lawson}, \citenamefont {Linden}, \citenamefont {Gajjar}, \citenamefont {Weniger},\ and\ \citenamefont {Safdi}}]{Foster:2022fxn}%
  \BibitemOpen
  \bibfield  {author} {\bibinfo {author} {\bibfnamefont {J.~W.}\ \bibnamefont {Foster}}, \bibinfo {author} {\bibfnamefont {S.~J.}\ \bibnamefont {Witte}}, \bibinfo {author} {\bibfnamefont {M.}~\bibnamefont {Lawson}}, \bibinfo {author} {\bibfnamefont {T.}~\bibnamefont {Linden}}, \bibinfo {author} {\bibfnamefont {V.}~\bibnamefont {Gajjar}}, \bibinfo {author} {\bibfnamefont {C.}~\bibnamefont {Weniger}}, \ and\ \bibinfo {author} {\bibfnamefont {B.~R.}\ \bibnamefont {Safdi}},\ }\href {\doibase 10.1103/PhysRevLett.129.251102} {\bibfield  {journal} {\bibinfo  {journal} {Phys. Rev. Lett.}\ }\textbf {\bibinfo {volume} {129}},\ \bibinfo {pages} {251102} (\bibinfo {year} {2022})},\ \Eprint {http://arxiv.org/abs/2202.08274} {arXiv:2202.08274 [astro-ph.CO]} \BibitemShut {NoStop}%
\bibitem [{\citenamefont {Battye}\ \emph {et~al.}(2023)\citenamefont {Battye}, \citenamefont {Keith}, \citenamefont {McDonald}, \citenamefont {Srinivasan}, \citenamefont {Stappers},\ and\ \citenamefont {Weltevrede}}]{Battye:2023oac}%
  \BibitemOpen
  \bibfield  {author} {\bibinfo {author} {\bibfnamefont {R.~A.}\ \bibnamefont {Battye}}, \bibinfo {author} {\bibfnamefont {M.~J.}\ \bibnamefont {Keith}}, \bibinfo {author} {\bibfnamefont {J.~I.}\ \bibnamefont {McDonald}}, \bibinfo {author} {\bibfnamefont {S.}~\bibnamefont {Srinivasan}}, \bibinfo {author} {\bibfnamefont {B.~W.}\ \bibnamefont {Stappers}}, \ and\ \bibinfo {author} {\bibfnamefont {P.}~\bibnamefont {Weltevrede}},\ }\href {\doibase 10.1103/PhysRevD.108.063001} {\bibfield  {journal} {\bibinfo  {journal} {Phys. Rev. D}\ }\textbf {\bibinfo {volume} {108}},\ \bibinfo {pages} {063001} (\bibinfo {year} {2023})},\ \Eprint {http://arxiv.org/abs/2303.11792} {arXiv:2303.11792 [astro-ph.CO]} \BibitemShut {NoStop}%
\bibitem [{\citenamefont {Buschmann}\ \emph {et~al.}(2021)\citenamefont {Buschmann}, \citenamefont {Co}, \citenamefont {Dessert},\ and\ \citenamefont {Safdi}}]{Buschmann:2019pfp}%
  \BibitemOpen
  \bibfield  {author} {\bibinfo {author} {\bibfnamefont {M.}~\bibnamefont {Buschmann}}, \bibinfo {author} {\bibfnamefont {R.~T.}\ \bibnamefont {Co}}, \bibinfo {author} {\bibfnamefont {C.}~\bibnamefont {Dessert}}, \ and\ \bibinfo {author} {\bibfnamefont {B.~R.}\ \bibnamefont {Safdi}},\ }\href {\doibase 10.1103/PhysRevLett.126.021102} {\bibfield  {journal} {\bibinfo  {journal} {Phys. Rev. Lett.}\ }\textbf {\bibinfo {volume} {126}},\ \bibinfo {pages} {021102} (\bibinfo {year} {2021})},\ \Eprint {http://arxiv.org/abs/1910.04164} {arXiv:1910.04164 [hep-ph]} \BibitemShut {NoStop}%
\bibitem [{\citenamefont {Leroy}\ \emph {et~al.}(2020)\citenamefont {Leroy}, \citenamefont {Chianese}, \citenamefont {Edwards},\ and\ \citenamefont {Weniger}}]{Leroy:2019ghm}%
  \BibitemOpen
  \bibfield  {author} {\bibinfo {author} {\bibfnamefont {M.}~\bibnamefont {Leroy}}, \bibinfo {author} {\bibfnamefont {M.}~\bibnamefont {Chianese}}, \bibinfo {author} {\bibfnamefont {T.~D.~P.}\ \bibnamefont {Edwards}}, \ and\ \bibinfo {author} {\bibfnamefont {C.}~\bibnamefont {Weniger}},\ }\href {\doibase 10.1103/PhysRevD.101.123003} {\bibfield  {journal} {\bibinfo  {journal} {Phys. Rev. D}\ }\textbf {\bibinfo {volume} {101}},\ \bibinfo {pages} {123003} (\bibinfo {year} {2020})},\ \Eprint {http://arxiv.org/abs/1912.08815} {arXiv:1912.08815 [hep-ph]} \BibitemShut {NoStop}%
\bibitem [{\citenamefont {Witte}\ \emph {et~al.}(2021)\citenamefont {Witte}, \citenamefont {Noordhuis}, \citenamefont {Edwards},\ and\ \citenamefont {Weniger}}]{Witte:2021arp}%
  \BibitemOpen
  \bibfield  {author} {\bibinfo {author} {\bibfnamefont {S.~J.}\ \bibnamefont {Witte}}, \bibinfo {author} {\bibfnamefont {D.}~\bibnamefont {Noordhuis}}, \bibinfo {author} {\bibfnamefont {T.~D.~P.}\ \bibnamefont {Edwards}}, \ and\ \bibinfo {author} {\bibfnamefont {C.}~\bibnamefont {Weniger}},\ }\href {\doibase 10.1103/PhysRevD.104.103030} {\bibfield  {journal} {\bibinfo  {journal} {Phys. Rev. D}\ }\textbf {\bibinfo {volume} {104}},\ \bibinfo {pages} {103030} (\bibinfo {year} {2021})},\ \Eprint {http://arxiv.org/abs/2104.07670} {arXiv:2104.07670 [hep-ph]} \BibitemShut {NoStop}%
\bibitem [{\citenamefont {Battye}\ \emph {et~al.}(2021)\citenamefont {Battye}, \citenamefont {Garbrecht}, \citenamefont {McDonald},\ and\ \citenamefont {Srinivasan}}]{Battye:2021xvt}%
  \BibitemOpen
  \bibfield  {author} {\bibinfo {author} {\bibfnamefont {R.~A.}\ \bibnamefont {Battye}}, \bibinfo {author} {\bibfnamefont {B.}~\bibnamefont {Garbrecht}}, \bibinfo {author} {\bibfnamefont {J.~I.}\ \bibnamefont {McDonald}}, \ and\ \bibinfo {author} {\bibfnamefont {S.}~\bibnamefont {Srinivasan}},\ }\href {\doibase 10.1007/JHEP09(2021)105} {\bibfield  {journal} {\bibinfo  {journal} {JHEP}\ }\textbf {\bibinfo {volume} {09}},\ \bibinfo {pages} {105} (\bibinfo {year} {2021})},\ \Eprint {http://arxiv.org/abs/2104.08290} {arXiv:2104.08290 [hep-ph]} \BibitemShut {NoStop}%
\bibitem [{\citenamefont {McDonald}\ and\ \citenamefont {Witte}(2023)}]{McDonald:2023shx}%
  \BibitemOpen
  \bibfield  {author} {\bibinfo {author} {\bibfnamefont {J.~I.}\ \bibnamefont {McDonald}}\ and\ \bibinfo {author} {\bibfnamefont {S.~J.}\ \bibnamefont {Witte}},\ }\href {\doibase 10.1103/PhysRevD.108.103021} {\bibfield  {journal} {\bibinfo  {journal} {Phys. Rev. D}\ }\textbf {\bibinfo {volume} {108}},\ \bibinfo {pages} {103021} (\bibinfo {year} {2023})},\ \Eprint {http://arxiv.org/abs/2309.08655} {arXiv:2309.08655 [hep-ph]} \BibitemShut {NoStop}%
\bibitem [{\citenamefont {Tjemsland}\ \emph {et~al.}(2024)\citenamefont {Tjemsland}, \citenamefont {McDonald},\ and\ \citenamefont {Witte}}]{Tjemsland:2023vvc}%
  \BibitemOpen
  \bibfield  {author} {\bibinfo {author} {\bibfnamefont {J.}~\bibnamefont {Tjemsland}}, \bibinfo {author} {\bibfnamefont {J.}~\bibnamefont {McDonald}}, \ and\ \bibinfo {author} {\bibfnamefont {S.~J.}\ \bibnamefont {Witte}},\ }\href {\doibase 10.1103/PhysRevD.109.023015} {\bibfield  {journal} {\bibinfo  {journal} {Phys. Rev. D}\ }\textbf {\bibinfo {volume} {109}},\ \bibinfo {pages} {023015} (\bibinfo {year} {2024})},\ \Eprint {http://arxiv.org/abs/2310.18403} {arXiv:2310.18403 [hep-ph]} \BibitemShut {NoStop}%
\bibitem [{\citenamefont {McDonald}\ \emph {et~al.}(2023)\citenamefont {McDonald}, \citenamefont {Garbrecht},\ and\ \citenamefont {Millington}}]{McDonald:2023ohd}%
  \BibitemOpen
  \bibfield  {author} {\bibinfo {author} {\bibfnamefont {J.~I.}\ \bibnamefont {McDonald}}, \bibinfo {author} {\bibfnamefont {B.}~\bibnamefont {Garbrecht}}, \ and\ \bibinfo {author} {\bibfnamefont {P.}~\bibnamefont {Millington}},\ }\href {\doibase 10.1088/1475-7516/2023/12/031} {\bibfield  {journal} {\bibinfo  {journal} {JCAP}\ }\textbf {\bibinfo {volume} {12}},\ \bibinfo {pages} {031} (\bibinfo {year} {2023})},\ \Eprint {http://arxiv.org/abs/2307.11812} {arXiv:2307.11812 [hep-ph]} \BibitemShut {NoStop}%
\bibitem [{\citenamefont {Gin\'es}\ \emph {et~al.}(2024)\citenamefont {Gin\'es}, \citenamefont {Noordhuis}, \citenamefont {Weniger},\ and\ \citenamefont {Witte}}]{Gines:2024ekm}%
  \BibitemOpen
  \bibfield  {author} {\bibinfo {author} {\bibfnamefont {E.~U.}\ \bibnamefont {Gin\'es}}, \bibinfo {author} {\bibfnamefont {D.}~\bibnamefont {Noordhuis}}, \bibinfo {author} {\bibfnamefont {C.}~\bibnamefont {Weniger}}, \ and\ \bibinfo {author} {\bibfnamefont {S.~J.}\ \bibnamefont {Witte}},\ }\href@noop {} {\  (\bibinfo {year} {2024})},\ \Eprint {http://arxiv.org/abs/2405.08865} {arXiv:2405.08865 [hep-ph]} \BibitemShut {NoStop}%
\bibitem [{\citenamefont {Ito}\ \emph {et~al.}(2024)\citenamefont {Ito}, \citenamefont {Kohri},\ and\ \citenamefont {Nakayama}}]{Ito:2023fcr}%
  \BibitemOpen
  \bibfield  {author} {\bibinfo {author} {\bibfnamefont {A.}~\bibnamefont {Ito}}, \bibinfo {author} {\bibfnamefont {K.}~\bibnamefont {Kohri}}, \ and\ \bibinfo {author} {\bibfnamefont {K.}~\bibnamefont {Nakayama}},\ }\href {\doibase 10.1103/PhysRevD.109.063026} {\bibfield  {journal} {\bibinfo  {journal} {Phys. Rev. D}\ }\textbf {\bibinfo {volume} {109}},\ \bibinfo {pages} {063026} (\bibinfo {year} {2024})},\ \Eprint {http://arxiv.org/abs/2305.13984} {arXiv:2305.13984 [gr-qc]} \BibitemShut {NoStop}%
\bibitem [{\citenamefont {Dandoy}\ \emph {et~al.}(2024)\citenamefont {Dandoy}, \citenamefont {Bert\'olez-Mart\'\i{}nez},\ and\ \citenamefont {Costa}}]{Dandoy:2024oqg}%
  \BibitemOpen
  \bibfield  {author} {\bibinfo {author} {\bibfnamefont {V.}~\bibnamefont {Dandoy}}, \bibinfo {author} {\bibfnamefont {T.}~\bibnamefont {Bert\'olez-Mart\'\i{}nez}}, \ and\ \bibinfo {author} {\bibfnamefont {F.}~\bibnamefont {Costa}},\ }\href@noop {} {\  (\bibinfo {year} {2024})},\ \Eprint {http://arxiv.org/abs/2402.14092} {arXiv:2402.14092 [gr-qc]} \BibitemShut {NoStop}%
\bibitem [{\citenamefont {Beskin}\ \emph {et~al.}(1986)\citenamefont {Beskin}, \citenamefont {Gurevich},\ and\ \citenamefont {Istomin}}]{VasiliiSBeskin1986}%
  \BibitemOpen
  \bibfield  {author} {\bibinfo {author} {\bibfnamefont {V.~S.}\ \bibnamefont {Beskin}}, \bibinfo {author} {\bibfnamefont {A.~V.}\ \bibnamefont {Gurevich}}, \ and\ \bibinfo {author} {\bibfnamefont {Y.~N.}\ \bibnamefont {Istomin}},\ }\href {\doibase 10.1070/PU1986v029n10ABEH003528} {\bibfield  {journal} {\bibinfo  {journal} {Soviet Physics Uspekhi}\ }\textbf {\bibinfo {volume} {29}},\ \bibinfo {pages} {946} (\bibinfo {year} {1986})}\BibitemShut {NoStop}%
\bibitem [{\citenamefont {Heisenberg}\ and\ \citenamefont {Euler}(1936)}]{Heisenberg:1936nmg}%
  \BibitemOpen
  \bibfield  {author} {\bibinfo {author} {\bibfnamefont {W.}~\bibnamefont {Heisenberg}}\ and\ \bibinfo {author} {\bibfnamefont {H.}~\bibnamefont {Euler}},\ }\href {\doibase 10.1007/BF01343663} {\bibfield  {journal} {\bibinfo  {journal} {Z. Phys.}\ }\textbf {\bibinfo {volume} {98}},\ \bibinfo {pages} {714} (\bibinfo {year} {1936})},\ \Eprint {http://arxiv.org/abs/physics/0605038} {arXiv:physics/0605038} \BibitemShut {NoStop}%
\bibitem [{\citenamefont {Adler}(1971)}]{Adler:1971wn}%
  \BibitemOpen
  \bibfield  {author} {\bibinfo {author} {\bibfnamefont {S.~L.}\ \bibnamefont {Adler}},\ }\href {\doibase 10.1016/0003-4916(71)90154-0} {\bibfield  {journal} {\bibinfo  {journal} {Annals Phys.}\ }\textbf {\bibinfo {volume} {67}},\ \bibinfo {pages} {599} (\bibinfo {year} {1971})}\BibitemShut {NoStop}%
\bibitem [{\citenamefont {Gedalin}\ \emph {et~al.}(1998)\citenamefont {Gedalin}, \citenamefont {Melrose},\ and\ \citenamefont {Gruman}}]{PhysRevE.57.3399}%
  \BibitemOpen
  \bibfield  {author} {\bibinfo {author} {\bibfnamefont {M.}~\bibnamefont {Gedalin}}, \bibinfo {author} {\bibfnamefont {D.~B.}\ \bibnamefont {Melrose}}, \ and\ \bibinfo {author} {\bibfnamefont {E.}~\bibnamefont {Gruman}},\ }\href {\doibase 10.1103/PhysRevE.57.3399} {\bibfield  {journal} {\bibinfo  {journal} {Phys. Rev. E}\ }\textbf {\bibinfo {volume} {57}},\ \bibinfo {pages} {3399} (\bibinfo {year} {1998})}\BibitemShut {NoStop}%
\bibitem [{\citenamefont {Goldreich}\ and\ \citenamefont {Julian}(1969)}]{Goldreich:1969sb}%
  \BibitemOpen
  \bibfield  {author} {\bibinfo {author} {\bibfnamefont {P.}~\bibnamefont {Goldreich}}\ and\ \bibinfo {author} {\bibfnamefont {W.~H.}\ \bibnamefont {Julian}},\ }\href {\doibase 10.1086/150119} {\bibfield  {journal} {\bibinfo  {journal} {Astrophys. J.}\ }\textbf {\bibinfo {volume} {157}},\ \bibinfo {pages} {869} (\bibinfo {year} {1969})}\BibitemShut {NoStop}%
\bibitem [{\citenamefont {Hare}\ \emph {et~al.}(2024)\citenamefont {Hare}, \citenamefont {Pavlov}, \citenamefont {Posselt}, \citenamefont {Kargaltsev}, \citenamefont {Temim},\ and\ \citenamefont {Chen}}]{Hare:2024lcv}%
  \BibitemOpen
  \bibfield  {author} {\bibinfo {author} {\bibfnamefont {J.}~\bibnamefont {Hare}}, \bibinfo {author} {\bibfnamefont {G.~G.}\ \bibnamefont {Pavlov}}, \bibinfo {author} {\bibfnamefont {B.}~\bibnamefont {Posselt}}, \bibinfo {author} {\bibfnamefont {O.}~\bibnamefont {Kargaltsev}}, \bibinfo {author} {\bibfnamefont {T.}~\bibnamefont {Temim}}, \ and\ \bibinfo {author} {\bibfnamefont {S.}~\bibnamefont {Chen}},\ }\href@noop {} {\  (\bibinfo {year} {2024})},\ \Eprint {http://arxiv.org/abs/2405.03947} {arXiv:2405.03947 [astro-ph.HE]} \BibitemShut {NoStop}%
\bibitem [{\citenamefont {Dessert}\ \emph {et~al.}(2020)\citenamefont {Dessert}, \citenamefont {Foster},\ and\ \citenamefont {Safdi}}]{Dessert:2019dos}%
  \BibitemOpen
  \bibfield  {author} {\bibinfo {author} {\bibfnamefont {C.}~\bibnamefont {Dessert}}, \bibinfo {author} {\bibfnamefont {J.~W.}\ \bibnamefont {Foster}}, \ and\ \bibinfo {author} {\bibfnamefont {B.~R.}\ \bibnamefont {Safdi}},\ }\href {\doibase 10.3847/1538-4357/abb4ea} {\bibfield  {journal} {\bibinfo  {journal} {Astrophys. J.}\ }\textbf {\bibinfo {volume} {904}},\ \bibinfo {pages} {42} (\bibinfo {year} {2020})},\ \Eprint {http://arxiv.org/abs/1910.02956} {arXiv:1910.02956 [astro-ph.HE]} \BibitemShut {NoStop}%
\bibitem [{\citenamefont {Fruscione}\ \emph {et~al.}(2006)\citenamefont {Fruscione}, \citenamefont {McDowell}, \citenamefont {Allen}, \citenamefont {Brickhouse}, \citenamefont {Burke}, \citenamefont {Davis}, \citenamefont {Durham}, \citenamefont {Elvis}, \citenamefont {Galle}, \citenamefont {Harris}, \citenamefont {Huenemoerder}, \citenamefont {Houck}, \citenamefont {Ishibashi}, \citenamefont {Karovska}, \citenamefont {Nicastro}, \citenamefont {Noble}, \citenamefont {Nowak}, \citenamefont {Primini}, \citenamefont {Siemiginowska}, \citenamefont {Smith},\ and\ \citenamefont {Wise}}]{10.1117/12.671760}%
  \BibitemOpen
  \bibfield  {author} {\bibinfo {author} {\bibfnamefont {A.}~\bibnamefont {Fruscione}}, \bibinfo {author} {\bibfnamefont {J.~C.}\ \bibnamefont {McDowell}}, \bibinfo {author} {\bibfnamefont {G.~E.}\ \bibnamefont {Allen}}, \bibinfo {author} {\bibfnamefont {N.~S.}\ \bibnamefont {Brickhouse}}, \bibinfo {author} {\bibfnamefont {D.~J.}\ \bibnamefont {Burke}}, \bibinfo {author} {\bibfnamefont {J.~E.}\ \bibnamefont {Davis}}, \bibinfo {author} {\bibfnamefont {N.}~\bibnamefont {Durham}}, \bibinfo {author} {\bibfnamefont {M.}~\bibnamefont {Elvis}}, \bibinfo {author} {\bibfnamefont {E.~C.}\ \bibnamefont {Galle}}, \bibinfo {author} {\bibfnamefont {D.~E.}\ \bibnamefont {Harris}}, \bibinfo {author} {\bibfnamefont {D.~P.}\ \bibnamefont {Huenemoerder}}, \bibinfo {author} {\bibfnamefont {J.~C.}\ \bibnamefont {Houck}}, \bibinfo {author} {\bibfnamefont {B.}~\bibnamefont {Ishibashi}}, \bibinfo {author} {\bibfnamefont {M.}~\bibnamefont {Karovska}}, \bibinfo {author} {\bibfnamefont {F.}~\bibnamefont {Nicastro}}, \bibinfo {author}
  {\bibfnamefont {M.~S.}\ \bibnamefont {Noble}}, \bibinfo {author} {\bibfnamefont {M.~A.}\ \bibnamefont {Nowak}}, \bibinfo {author} {\bibfnamefont {F.~A.}\ \bibnamefont {Primini}}, \bibinfo {author} {\bibfnamefont {A.}~\bibnamefont {Siemiginowska}}, \bibinfo {author} {\bibfnamefont {R.~K.}\ \bibnamefont {Smith}}, \ and\ \bibinfo {author} {\bibfnamefont {M.}~\bibnamefont {Wise}},\ }in\ \href {\doibase 10.1117/12.671760} {\emph {\bibinfo {booktitle} {Observatory Operations: Strategies, Processes, and Systems}}},\ Vol.\ \bibinfo {volume} {6270},\ \bibinfo {editor} {edited by\ \bibinfo {editor} {\bibfnamefont {D.~R.}\ \bibnamefont {Silva}}\ and\ \bibinfo {editor} {\bibfnamefont {R.~E.}\ \bibnamefont {Doxsey}}},\ \bibinfo {organization} {International Society for Optics and Photonics}\ (\bibinfo  {publisher} {SPIE},\ \bibinfo {year} {2006})\ p.\ \bibinfo {pages} {62701V}\BibitemShut {NoStop}%
\bibitem [{\citenamefont {{Harrison}}\ \emph {et~al.}(2013)\citenamefont {{Harrison}}, \citenamefont {{Craig}}, \citenamefont {{Christensen}}, \citenamefont {{Hailey}}, \citenamefont {{Zhang}}, \citenamefont {{Boggs}}, \citenamefont {{Stern}}, \citenamefont {{Cook}}, \citenamefont {{Forster}}, \citenamefont {{Giommi}}, \citenamefont {{Grefenstette}}, \citenamefont {{Kim}}, \citenamefont {{Kitaguchi}}, \citenamefont {{Koglin}}, \citenamefont {{Madsen}}, \citenamefont {{Mao}}, \citenamefont {{Miyasaka}}, \citenamefont {{Mori}}, \citenamefont {{Perri}}, \citenamefont {{Pivovaroff}}, \citenamefont {{Puccetti}}, \citenamefont {{Rana}}, \citenamefont {{Westergaard}}, \citenamefont {{Willis}}, \citenamefont {{Zoglauer}}, \citenamefont {{An}}, \citenamefont {{Bachetti}}, \citenamefont {{Barri{\`e}re}}, \citenamefont {{Bellm}}, \citenamefont {{Bhalerao}}, \citenamefont {{Brejnholt}}, \citenamefont {{Fuerst}}, \citenamefont {{Liebe}}, \citenamefont {{Markwardt}}, \citenamefont {{Nynka}}, \citenamefont {{Vogel}},
  \citenamefont {{Walton}}, \citenamefont {{Wik}}, \citenamefont {{Alexander}}, \citenamefont {{Cominsky}}, \citenamefont {{Hornschemeier}}, \citenamefont {{Hornstrup}}, \citenamefont {{Kaspi}}, \citenamefont {{Madejski}}, \citenamefont {{Matt}}, \citenamefont {{Molendi}}, \citenamefont {{Smith}}, \citenamefont {{Tomsick}}, \citenamefont {{Ajello}}, \citenamefont {{Ballantyne}}, \citenamefont {{Balokovi{\'c}}}, \citenamefont {{Barret}}, \citenamefont {{Bauer}}, \citenamefont {{Blandford}}, \citenamefont {{Brandt}}, \citenamefont {{Brenneman}}, \citenamefont {{Chiang}}, \citenamefont {{Chakrabarty}}, \citenamefont {{Chenevez}}, \citenamefont {{Comastri}}, \citenamefont {{Dufour}}, \citenamefont {{Elvis}}, \citenamefont {{Fabian}}, \citenamefont {{Farrah}}, \citenamefont {{Fryer}}, \citenamefont {{Gotthelf}}, \citenamefont {{Grindlay}}, \citenamefont {{Helfand}}, \citenamefont {{Krivonos}}, \citenamefont {{Meier}}, \citenamefont {{Miller}}, \citenamefont {{Natalucci}}, \citenamefont {{Ogle}}, \citenamefont
  {{Ofek}}, \citenamefont {{Ptak}}, \citenamefont {{Reynolds}}, \citenamefont {{Rigby}}, \citenamefont {{Tagliaferri}}, \citenamefont {{Thorsett}}, \citenamefont {{Treister}},\ and\ \citenamefont {{Urry}}}]{2013ApJ...770..103H}%
  \BibitemOpen
  \bibfield  {author} {\bibinfo {author} {\bibfnamefont {F.~A.}\ \bibnamefont {{Harrison}}}, \bibinfo {author} {\bibfnamefont {W.~W.}\ \bibnamefont {{Craig}}}, \bibinfo {author} {\bibfnamefont {F.~E.}\ \bibnamefont {{Christensen}}}, \bibinfo {author} {\bibfnamefont {C.~J.}\ \bibnamefont {{Hailey}}}, \bibinfo {author} {\bibfnamefont {W.~W.}\ \bibnamefont {{Zhang}}}, \bibinfo {author} {\bibfnamefont {S.~E.}\ \bibnamefont {{Boggs}}}, \bibinfo {author} {\bibfnamefont {D.}~\bibnamefont {{Stern}}}, \bibinfo {author} {\bibfnamefont {W.~R.}\ \bibnamefont {{Cook}}}, \bibinfo {author} {\bibfnamefont {K.}~\bibnamefont {{Forster}}}, \bibinfo {author} {\bibfnamefont {P.}~\bibnamefont {{Giommi}}}, \bibinfo {author} {\bibfnamefont {B.~W.}\ \bibnamefont {{Grefenstette}}}, \bibinfo {author} {\bibfnamefont {Y.}~\bibnamefont {{Kim}}}, \bibinfo {author} {\bibfnamefont {T.}~\bibnamefont {{Kitaguchi}}}, \bibinfo {author} {\bibfnamefont {J.~E.}\ \bibnamefont {{Koglin}}}, \bibinfo {author} {\bibfnamefont {K.~K.}\ \bibnamefont
  {{Madsen}}}, \bibinfo {author} {\bibfnamefont {P.~H.}\ \bibnamefont {{Mao}}}, \bibinfo {author} {\bibfnamefont {H.}~\bibnamefont {{Miyasaka}}}, \bibinfo {author} {\bibfnamefont {K.}~\bibnamefont {{Mori}}}, \bibinfo {author} {\bibfnamefont {M.}~\bibnamefont {{Perri}}}, \bibinfo {author} {\bibfnamefont {M.~J.}\ \bibnamefont {{Pivovaroff}}}, \bibinfo {author} {\bibfnamefont {S.}~\bibnamefont {{Puccetti}}}, \bibinfo {author} {\bibfnamefont {V.~R.}\ \bibnamefont {{Rana}}}, \bibinfo {author} {\bibfnamefont {N.~J.}\ \bibnamefont {{Westergaard}}}, \bibinfo {author} {\bibfnamefont {J.}~\bibnamefont {{Willis}}}, \bibinfo {author} {\bibfnamefont {A.}~\bibnamefont {{Zoglauer}}}, \bibinfo {author} {\bibfnamefont {H.}~\bibnamefont {{An}}}, \bibinfo {author} {\bibfnamefont {M.}~\bibnamefont {{Bachetti}}}, \bibinfo {author} {\bibfnamefont {N.~M.}\ \bibnamefont {{Barri{\`e}re}}}, \bibinfo {author} {\bibfnamefont {E.~C.}\ \bibnamefont {{Bellm}}}, \bibinfo {author} {\bibfnamefont {V.}~\bibnamefont {{Bhalerao}}}, \bibinfo
  {author} {\bibfnamefont {N.~F.}\ \bibnamefont {{Brejnholt}}}, \bibinfo {author} {\bibfnamefont {F.}~\bibnamefont {{Fuerst}}}, \bibinfo {author} {\bibfnamefont {C.~C.}\ \bibnamefont {{Liebe}}}, \bibinfo {author} {\bibfnamefont {C.~B.}\ \bibnamefont {{Markwardt}}}, \bibinfo {author} {\bibfnamefont {M.}~\bibnamefont {{Nynka}}}, \bibinfo {author} {\bibfnamefont {J.~K.}\ \bibnamefont {{Vogel}}}, \bibinfo {author} {\bibfnamefont {D.~J.}\ \bibnamefont {{Walton}}}, \bibinfo {author} {\bibfnamefont {D.~R.}\ \bibnamefont {{Wik}}}, \bibinfo {author} {\bibfnamefont {D.~M.}\ \bibnamefont {{Alexander}}}, \bibinfo {author} {\bibfnamefont {L.~R.}\ \bibnamefont {{Cominsky}}}, \bibinfo {author} {\bibfnamefont {A.~E.}\ \bibnamefont {{Hornschemeier}}}, \bibinfo {author} {\bibfnamefont {A.}~\bibnamefont {{Hornstrup}}}, \bibinfo {author} {\bibfnamefont {V.~M.}\ \bibnamefont {{Kaspi}}}, \bibinfo {author} {\bibfnamefont {G.~M.}\ \bibnamefont {{Madejski}}}, \bibinfo {author} {\bibfnamefont {G.}~\bibnamefont {{Matt}}}, \bibinfo
  {author} {\bibfnamefont {S.}~\bibnamefont {{Molendi}}}, \bibinfo {author} {\bibfnamefont {D.~M.}\ \bibnamefont {{Smith}}}, \bibinfo {author} {\bibfnamefont {J.~A.}\ \bibnamefont {{Tomsick}}}, \bibinfo {author} {\bibfnamefont {M.}~\bibnamefont {{Ajello}}}, \bibinfo {author} {\bibfnamefont {D.~R.}\ \bibnamefont {{Ballantyne}}}, \bibinfo {author} {\bibfnamefont {M.}~\bibnamefont {{Balokovi{\'c}}}}, \bibinfo {author} {\bibfnamefont {D.}~\bibnamefont {{Barret}}}, \bibinfo {author} {\bibfnamefont {F.~E.}\ \bibnamefont {{Bauer}}}, \bibinfo {author} {\bibfnamefont {R.~D.}\ \bibnamefont {{Blandford}}}, \bibinfo {author} {\bibfnamefont {W.~N.}\ \bibnamefont {{Brandt}}}, \bibinfo {author} {\bibfnamefont {L.~W.}\ \bibnamefont {{Brenneman}}}, \bibinfo {author} {\bibfnamefont {J.}~\bibnamefont {{Chiang}}}, \bibinfo {author} {\bibfnamefont {D.}~\bibnamefont {{Chakrabarty}}}, \bibinfo {author} {\bibfnamefont {J.}~\bibnamefont {{Chenevez}}}, \bibinfo {author} {\bibfnamefont {A.}~\bibnamefont {{Comastri}}}, \bibinfo {author}
  {\bibfnamefont {F.}~\bibnamefont {{Dufour}}}, \bibinfo {author} {\bibfnamefont {M.}~\bibnamefont {{Elvis}}}, \bibinfo {author} {\bibfnamefont {A.~C.}\ \bibnamefont {{Fabian}}}, \bibinfo {author} {\bibfnamefont {D.}~\bibnamefont {{Farrah}}}, \bibinfo {author} {\bibfnamefont {C.~L.}\ \bibnamefont {{Fryer}}}, \bibinfo {author} {\bibfnamefont {E.~V.}\ \bibnamefont {{Gotthelf}}}, \bibinfo {author} {\bibfnamefont {J.~E.}\ \bibnamefont {{Grindlay}}}, \bibinfo {author} {\bibfnamefont {D.~J.}\ \bibnamefont {{Helfand}}}, \bibinfo {author} {\bibfnamefont {R.}~\bibnamefont {{Krivonos}}}, \bibinfo {author} {\bibfnamefont {D.~L.}\ \bibnamefont {{Meier}}}, \bibinfo {author} {\bibfnamefont {J.~M.}\ \bibnamefont {{Miller}}}, \bibinfo {author} {\bibfnamefont {L.}~\bibnamefont {{Natalucci}}}, \bibinfo {author} {\bibfnamefont {P.}~\bibnamefont {{Ogle}}}, \bibinfo {author} {\bibfnamefont {E.~O.}\ \bibnamefont {{Ofek}}}, \bibinfo {author} {\bibfnamefont {A.}~\bibnamefont {{Ptak}}}, \bibinfo {author} {\bibfnamefont {S.~P.}\
  \bibnamefont {{Reynolds}}}, \bibinfo {author} {\bibfnamefont {J.~R.}\ \bibnamefont {{Rigby}}}, \bibinfo {author} {\bibfnamefont {G.}~\bibnamefont {{Tagliaferri}}}, \bibinfo {author} {\bibfnamefont {S.~E.}\ \bibnamefont {{Thorsett}}}, \bibinfo {author} {\bibfnamefont {E.}~\bibnamefont {{Treister}}}, \ and\ \bibinfo {author} {\bibfnamefont {C.~M.}\ \bibnamefont {{Urry}}},\ }\href {\doibase 10.1088/0004-637X/770/2/103} {\bibfield  {journal} {\bibinfo  {journal} {\apj}\ }\textbf {\bibinfo {volume} {770}},\ \bibinfo {eid} {103} (\bibinfo {year} {2013})},\ \Eprint {http://arxiv.org/abs/1301.7307} {arXiv:1301.7307 [astro-ph.IM]} \BibitemShut {NoStop}%
\bibitem [{NuS()}]{NuStarHEASARC}%
  \BibitemOpen
  \href {https://heasarc.gsfc.nasa.gov/cgi-bin/W3Browse/w3query.pl?Radius=Default&Radius_unit=arcmin&Coordinates=Equatorial&Equinox=2000&ResultMax=0&NR=CheckCaches%2FGRB%2FSIMBAD%2BSesame%2FNED&Entry=RX+J1856%2E6%2D3754&tablehead=name=heasarc_numaster&description=NuSTAR%20Master%20Catalog&url=https://heasarc.gsfc.nasa.gov/W3Browse/nustar/numaster.html&archive=Y&radius=10&mission=NUSTAR&priority=5&tabletype=Observation#} {\enquote {\bibinfo {title} {Nustar obsid 30801015002},}\ }\BibitemShut {NoStop}%
\bibitem [{\citenamefont {Abdo}\ \emph {et~al.}(2010)\citenamefont {Abdo} \emph {et~al.}}]{Fermi-LAT:2010mou}%
  \BibitemOpen
  \bibfield  {author} {\bibinfo {author} {\bibfnamefont {A.~A.}\ \bibnamefont {Abdo}} \emph {et~al.} (\bibinfo {collaboration} {Fermi-LAT}),\ }\href {\doibase 10.1088/0004-637X/720/1/272} {\bibfield  {journal} {\bibinfo  {journal} {Astrophys. J.}\ }\textbf {\bibinfo {volume} {720}},\ \bibinfo {pages} {272} (\bibinfo {year} {2010})},\ \Eprint {http://arxiv.org/abs/1007.1142} {arXiv:1007.1142 [astro-ph.HE]} \BibitemShut {NoStop}%
\bibitem [{\citenamefont {{Thompson}}(1996)}]{1996ASPC..105..307T}%
  \BibitemOpen
  \bibfield  {author} {\bibinfo {author} {\bibfnamefont {D.~J.}\ \bibnamefont {{Thompson}}},\ }in\ \href@noop {} {\emph {\bibinfo {booktitle} {IAU Colloq. 160: Pulsars: Problems and Progress}}},\ \bibinfo {series} {Astronomical Society of the Pacific Conference Series}, Vol.\ \bibinfo {volume} {105},\ \bibinfo {editor} {edited by\ \bibinfo {editor} {\bibfnamefont {S.}~\bibnamefont {{Johnston}}}, \bibinfo {editor} {\bibfnamefont {M.~A.}\ \bibnamefont {{Walker}}}, \ and\ \bibinfo {editor} {\bibfnamefont {M.}~\bibnamefont {{Bailes}}}}\ (\bibinfo {year} {1996})\ p.\ \bibinfo {pages} {307}\BibitemShut {NoStop}%
\bibitem [{\citenamefont {Meyer}\ \emph {et~al.}(2010)\citenamefont {Meyer}, \citenamefont {Horns},\ and\ \citenamefont {Zechlin}}]{Meyer:2010tta}%
  \BibitemOpen
  \bibfield  {author} {\bibinfo {author} {\bibfnamefont {M.}~\bibnamefont {Meyer}}, \bibinfo {author} {\bibfnamefont {D.}~\bibnamefont {Horns}}, \ and\ \bibinfo {author} {\bibfnamefont {H.~S.}\ \bibnamefont {Zechlin}},\ }\href {\doibase 10.1051/0004-6361/201014108} {\bibfield  {journal} {\bibinfo  {journal} {Astron. Astrophys.}\ }\textbf {\bibinfo {volume} {523}},\ \bibinfo {pages} {A2} (\bibinfo {year} {2010})},\ \Eprint {http://arxiv.org/abs/1008.4524} {arXiv:1008.4524 [astro-ph.HE]} \BibitemShut {NoStop}%
\bibitem [{\citenamefont {Hill}\ \emph {et~al.}(2018)\citenamefont {Hill}, \citenamefont {Masui},\ and\ \citenamefont {Scott}}]{Hill:2018trh}%
  \BibitemOpen
  \bibfield  {author} {\bibinfo {author} {\bibfnamefont {R.}~\bibnamefont {Hill}}, \bibinfo {author} {\bibfnamefont {K.~W.}\ \bibnamefont {Masui}}, \ and\ \bibinfo {author} {\bibfnamefont {D.}~\bibnamefont {Scott}},\ }\href {\doibase 10.1177/0003702818767133} {\bibfield  {journal} {\bibinfo  {journal} {Appl. Spectrosc.}\ }\textbf {\bibinfo {volume} {72}},\ \bibinfo {pages} {663} (\bibinfo {year} {2018})},\ \Eprint {http://arxiv.org/abs/1802.03694} {arXiv:1802.03694 [astro-ph.CO]} \BibitemShut {NoStop}%
\bibitem [{\citenamefont {Ejlli}\ \emph {et~al.}(2019)\citenamefont {Ejlli}, \citenamefont {Ejlli}, \citenamefont {Cruise}, \citenamefont {Pisano},\ and\ \citenamefont {Grote}}]{Ejlli:2019bqj}%
  \BibitemOpen
  \bibfield  {author} {\bibinfo {author} {\bibfnamefont {A.}~\bibnamefont {Ejlli}}, \bibinfo {author} {\bibfnamefont {D.}~\bibnamefont {Ejlli}}, \bibinfo {author} {\bibfnamefont {A.~M.}\ \bibnamefont {Cruise}}, \bibinfo {author} {\bibfnamefont {G.}~\bibnamefont {Pisano}}, \ and\ \bibinfo {author} {\bibfnamefont {H.}~\bibnamefont {Grote}},\ }\href {\doibase 10.1140/epjc/s10052-019-7542-5} {\bibfield  {journal} {\bibinfo  {journal} {Eur. Phys. J. C}\ }\textbf {\bibinfo {volume} {79}},\ \bibinfo {pages} {1032} (\bibinfo {year} {2019})},\ \Eprint {http://arxiv.org/abs/1908.00232} {arXiv:1908.00232 [gr-qc]} \BibitemShut {NoStop}%
\bibitem [{\citenamefont {Yeh}\ \emph {et~al.}(2022)\citenamefont {Yeh}, \citenamefont {Shelton}, \citenamefont {Olive},\ and\ \citenamefont {Fields}}]{Yeh:2022heq}%
  \BibitemOpen
  \bibfield  {author} {\bibinfo {author} {\bibfnamefont {T.-H.}\ \bibnamefont {Yeh}}, \bibinfo {author} {\bibfnamefont {J.}~\bibnamefont {Shelton}}, \bibinfo {author} {\bibfnamefont {K.~A.}\ \bibnamefont {Olive}}, \ and\ \bibinfo {author} {\bibfnamefont {B.~D.}\ \bibnamefont {Fields}},\ }\href {\doibase 10.1088/1475-7516/2022/10/046} {\bibfield  {journal} {\bibinfo  {journal} {JCAP}\ }\textbf {\bibinfo {volume} {10}},\ \bibinfo {pages} {046} (\bibinfo {year} {2022})},\ \Eprint {http://arxiv.org/abs/2207.13133} {arXiv:2207.13133 [astro-ph.CO]} \BibitemShut {NoStop}%
\bibitem [{\citenamefont {Domcke}\ and\ \citenamefont {Garcia-Cely}(2021)}]{Domcke:2020yzq}%
  \BibitemOpen
  \bibfield  {author} {\bibinfo {author} {\bibfnamefont {V.}~\bibnamefont {Domcke}}\ and\ \bibinfo {author} {\bibfnamefont {C.}~\bibnamefont {Garcia-Cely}},\ }\href {\doibase 10.1103/PhysRevLett.126.021104} {\bibfield  {journal} {\bibinfo  {journal} {Phys. Rev. Lett.}\ }\textbf {\bibinfo {volume} {126}},\ \bibinfo {pages} {021104} (\bibinfo {year} {2021})},\ \Eprint {http://arxiv.org/abs/2006.01161} {arXiv:2006.01161 [astro-ph.CO]} \BibitemShut {NoStop}%
\end{thebibliography}%

\end{document}